\documentclass[preprint2]{aastex62}

\usepackage{bm}
\usepackage{comment}
\usepackage{colortbl}

\newcommand{\red}{\textcolor{black}}
\newcommand{\redred}{\textcolor{black}}
\newcolumntype{a}{>{\columncolor{pink}}c}

\begin{document}

\title{Kojima-1Lb Is a Mildly Cold Neptune around the Brightest Microlensing Host Star}

\author[0000-0002-4909-5763]{A. Fukui}
\affiliation{Department of Earth and Planetary Science, Graduate School of Science, The University of Tokyo, 7-3-1 Hongo, Bunkyo-ku, Tokyo 113-0033, Japan}
\affiliation{Instituto de Astrof\'isica de Canarias, V\'ia L\'actea s/n, E-38205 La Laguna, Tenerife, Spain}

\author[0000-0002-5843-9433]{D. Suzuki}
\affiliation{Institute of Space and Astronautical Science, Japan Aerospace Exploration Agency (JAXA), 3-1-1 Yoshinodai, Chuo, Sagamihara, Kanagawa 252-5210, Japan}

\author{N. Koshimoto}
\affiliation{Department of Astronomy, Graduate School of Science, The University of Tokyo, 7-3-1 Hongo, Bunkyo-ku, Tokyo 113-0033, Japan}
\affiliation{National Astronomical Observatory of Japan, 2-21-1 Osawa, Mitaka, Tokyo 181-8588, Japan}
\affiliation{Code 667, NASA Goddard Space Flight Center, Greenbelt, MD 20771, USA}

\author{E. Bachelet}
\affiliation{Las Cumbres Observatory, 6740 Cortona Drive, Suite 102, Goleta, CA 93117, USA}

\author{T. Vanmunster}
\affiliation{Center for Backyard Astrophysics Belgium, Walhostraat 1A, B-3401 Landen, Belgium}

\author{D. Storey}
\affiliation{American Association of Variable Star Observers, 49 Bay State Road, Cambridge, MA 02138, USA}

\author{H. Maehara}
\affiliation{Okayama Branch Office, Subaru Telescope, National Astronomical Observatory of Japan, NINS, Kamogata, Asakuchi, Okayama 719-0232, Japan}

\author{K. Yanagisawa}
\affiliation{Division of Optical and Infrared Astronomy, National Astronomical Observatory of Japan, 2-21-1 Osawa, Mitaka-shi, Tokyo 181-8588, Japan}

\author{T. Yamada}
\affiliation{Department of Astrophysics and Atmospheric Sciences, Faculty of Science, Kyoto Sangyo University, 603-8555 Kyoto, Japan}

\author{A. Yonehara}
\affiliation{Department of Astrophysics and Atmospheric Sciences, Faculty of Science, Kyoto Sangyo University, 603-8555 Kyoto, Japan}

\author{T. Hirano}
\affiliation{Department of Earth and Planetary Sciences, Tokyo Institute of Technology, 2-12-1 Ookayama, Meguro-ku, Tokyo 152-8551, Japan}

\author{D. P. Bennett}
\affiliation{Code 667, NASA Goddard Space Flight Center, Greenbelt, MD 20771, USA}
\affiliation{Department of Astronomy, University of Maryland, College Park, MD 20742, USA}

\author{V. Bozza}
\affiliation{Dipartimento di Fisica ``E.R. Caianiello,'' Universit\'{a} di Salerno, Via Giovanni Paolo II 132, I-84084, Fisciano, Italy}
\affiliation{Istituto Nazionale di Fisica Nucleare, Sezione di Napoli, Via Cintia, I-80126 Napoli, Italy}

\author{D. Mawet}
\affiliation{Department of Astronomy, California Institute of Technology, Pasadena, CA 91125, USA}
\affiliation{Jet Propulsion Laboratory, California Institute of Technology, Pasadena, CA 91109, USA}

\author{M. T. Penny}
\affiliation{Department of Astronomy, The Ohio State University, 140 West 18th Avenue, Columbus, OH 43210, USA}

\author{S. Awiphan}
\affiliation{National Astronomical Research Institute of Thailand, 260, Moo 4, T. Donkaew, A. Mae Rim, Chiang Mai, 50180, Thailand}

\author{A. Oksanen}
\affiliation{American Association of Variable Star Observers, 49 Bay State Road, Cambridge, MA 02138, USA}
\affiliation{Hankasalmi Observatory, Hankasalmi, Finland}

\author{T. M. Heintz}
\affiliation{Institute for Astrophysical Research, Boston University, 725 Commonwealth Avenue, Boston, MA 02215, USA}
\affiliation{Department of Physics, Westminster College, New Wilmington, PA 16172, USA}

\author{T. E. Oberst}
\affiliation{Department of Physics, Westminster College, New Wilmington, PA 16172, USA}

\author{V. J. S. B\'{e}jar}
\affiliation{Instituto de Astrof\'isica de Canarias, V\'ia L\'actea s/n, E-38205 La Laguna, Tenerife, Spain}
\affiliation{Departamento de Astrof\'isica, Universidad de La Laguna (ULL), E-38206 La Laguna, Tenerife, Spain}

\author{N. Casasayas-Barris}
\affiliation{Instituto de Astrof\'isica de Canarias, V\'ia L\'actea s/n, E-38205 La Laguna, Tenerife, Spain}
\affiliation{Departamento de Astrof\'isica, Universidad de La Laguna (ULL), E-38206 La Laguna, Tenerife, Spain}

\author{G. Chen}
\affiliation{Key Laboratory of Planetary Sciences, Purple Mountain Observatory, Chinese Academy of Sciences, Nanjing 210008, China}
\affiliation{Instituto de Astrof\'isica de Canarias, V\'ia L\'actea s/n, E-38205 La Laguna, Tenerife, Spain}

\author{N. Crouzet}
\affiliation{European Space Agency, European Space Research and Technology Centre (ESTEC), Keplerlaan 1, 2201 AZ Noordwijk, The Netherlands}
\affiliation{Instituto de Astrof\'isica de Canarias, V\'ia L\'actea s/n, E-38205 La Laguna, Tenerife, Spain}
\affiliation{Departamento de Astrof\'isica, Universidad de La Laguna (ULL), E-38206 La Laguna, Tenerife, Spain}

\author{D. Hidalgo}
\affiliation{Instituto de Astrof\'isica de Canarias, V\'ia L\'actea s/n, E-38205 La Laguna, Tenerife, Spain}
\affiliation{Departamento de Astrof\'isica, Universidad de La Laguna (ULL), E-38206 La Laguna, Tenerife, Spain}

\author{P. Klagyivik}
\affiliation{Instituto de Astrof\'isica de Canarias, V\'ia L\'actea s/n, E-38205 La Laguna, Tenerife, Spain}
\affiliation{Departamento de Astrof\'isica, Universidad de La Laguna (ULL), E-38206 La Laguna, Tenerife, Spain}

\author{F. Murgas}
\affiliation{Instituto de Astrof\'isica de Canarias, V\'ia L\'actea s/n, E-38205 La Laguna, Tenerife, Spain}
\affiliation{Departamento de Astrof\'isica, Universidad de La Laguna (ULL), E-38206 La Laguna, Tenerife, Spain}

\author{N. Narita}
\affiliation{Astrobiology Center, National Institutes of Natural Sciences, 2-21-1 Osawa, Mitaka, Tokyo 181-8588, Japan}
\affiliation{Instituto de Astrof\'isica de Canarias, V\'ia L\'actea s/n, E-38205 La Laguna, Tenerife, Spain}
\affiliation{JST, PRESTO, 2-21-1 Osawa, Mitaka, Tokyo 181-8588, Japan}
\affiliation{National Astronomical Observatory of Japan, 2-21-1 Osawa, Mitaka, Tokyo 181-8588, Japan}

\author{E. Palle}
\affiliation{Instituto de Astrof\'isica de Canarias, V\'ia L\'actea s/n, E-38205 La Laguna, Tenerife, Spain}
\affiliation{Departamento de Astrof\'isica, Universidad de La Laguna (ULL), E-38206 La Laguna, Tenerife, Spain}

\author{H.  Parviainen}
\affiliation{Instituto de Astrof\'isica de Canarias, V\'ia L\'actea s/n, E-38205 La Laguna, Tenerife, Spain}
\affiliation{Departamento de Astrof\'isica, Universidad de La Laguna (ULL), E-38206 La Laguna, Tenerife, Spain}

\author{N. Watanabe}
\affiliation{SOKENDAI (The Graduate University of Advanced Studies), 2-21-1 Osawa, Mitaka, Tokyo 181-8588, Japan}
\affiliation{National Astronomical Observatory of Japan, 2-21-1 Osawa, Mitaka, Tokyo 181-8588, Japan}

\author{N. Kusakabe}
\affiliation{Astrobiology Center, National Institutes of Natural Sciences, 2-21-1 Osawa, Mitaka, Tokyo 181-8588, Japan}
\affiliation{National Astronomical Observatory of Japan, 2-21-1 Osawa, Mitaka, Tokyo 181-8588, Japan}

\author{M. Mori}
\affiliation{Department of Astronomy, Graduate School of Science, The University of Tokyo, 7-3-1 Hongo, Bunkyo-ku, Tokyo 113-0033, Japan}

\author{Y. Terada}
\affiliation{Department of Astronomy, Graduate School of Science, The University of Tokyo, 7-3-1 Hongo, Bunkyo-ku, Tokyo 113-0033, Japan}

\author{J. P. de Leon}
\affiliation{Department of Astronomy, Graduate School of Science, The University of Tokyo, 7-3-1 Hongo, Bunkyo-ku, Tokyo 113-0033, Japan}

\author{A. Hernandez}
\affiliation{Instituto de Astrof\'isica de Canarias, V\'ia L\'actea s/n, E-38205 La Laguna, Tenerife, Spain}
\affiliation{Departamento de Astrof\'isica, Universidad de La Laguna (ULL), E-38206 La Laguna, Tenerife, Spain}

\author{R. Luque}
\affiliation{Instituto de Astrof\'isica de Canarias, V\'ia L\'actea s/n, E-38205 La Laguna, Tenerife, Spain}
\affiliation{Departamento de Astrof\'isica, Universidad de La Laguna (ULL), E-38206 La Laguna, Tenerife, Spain}

\author{M. Monelli}
\affiliation{Instituto de Astrof\'isica de Canarias, V\'ia L\'actea s/n, E-38205 La Laguna, Tenerife, Spain}
\affiliation{Departamento de Astrof\'isica, Universidad de La Laguna (ULL), E-38206 La Laguna, Tenerife, Spain}

\author{P. Monta\~nes-Rodriguez}
\affiliation{Instituto de Astrof\'isica de Canarias, V\'ia L\'actea s/n, E-38205 La Laguna, Tenerife, Spain}
\affiliation{Departamento de Astrof\'isica, Universidad de La Laguna (ULL), E-38206 La Laguna, Tenerife, Spain}

\author{J. Prieto-Arranz}
\affiliation{Instituto de Astrof\'isica de Canarias, V\'ia L\'actea s/n, E-38205 La Laguna, Tenerife, Spain}
\affiliation{Departamento de Astrof\'isica, Universidad de La Laguna (ULL), E-38206 La Laguna, Tenerife, Spain}

\author{K. L. Murata}
\affiliation{Department of Physics, Tokyo Institute of Technology, 2-12-1 Ookayama, Meguro-ku, Tokyo 152-8551, Japan}

\author{S. Shugarov}
\affiliation{Sternberg Astronomical Institute, Moscow State University, Moscow 119991, Russia}
\affiliation{Astronomical Institute of the Slovak Academy of Sciences, Tatranska Lomnica, 05960, Slovakia}

\author{Y. Kubota}
\affiliation{Department of Astrophysics and Atmospheric Sciences, Faculty of Science, Kyoto Sangyo University, 603-8555 Kyoto, Japan}

\author{C. Otsuki}
\affiliation{Department of Astrophysics and Atmospheric Sciences, Faculty of Science, Kyoto Sangyo University, 603-8555 Kyoto, Japan}

\author{A. Shionoya}
\affiliation{Department of Astrophysics and Atmospheric Sciences, Faculty of Science, Kyoto Sangyo University, 603-8555 Kyoto, Japan}

\author{T. Nishiumi}
\affiliation{Department of Astrophysics and Atmospheric Sciences, Faculty of Science, Kyoto Sangyo University, 603-8555 Kyoto, Japan}
\affiliation{National Astronomical Observatory of Japan, 2-21-1 Osawa, Mitaka, Tokyo 181-8588, Japan}

\author{A. Nishide}
\affiliation{Department of Astrophysics and Atmospheric Sciences, Faculty of Science, Kyoto Sangyo University, 603-8555 Kyoto, Japan}

\author{M. Fukagawa}
\affiliation{National Astronomical Observatory of Japan, 2-21-2, Osawa, Mitaka, Tokyo 181-8588, Japan}

\author{K. Onodera}
\affiliation{Institute of Space and Astronautical Science, Japan Aerospace Exploration Agency (JAXA), 3-1-1 Yoshinodai, Chuo, Sagamihara, Kanagawa 252-5210, Japan}
\affiliation{Department of Space and Astronautical Science, SOKENDAI (The Graduate University for Advanced Studies), 3-1-1 Yoshinodai, Chuo-ku, Sagamihara, Kanagawa 252-5210, Japan}
\affiliation{Institut de Physique du Globe de Paris, 1 Rue Jussieu, F-75005 Paris, France}
\affiliation{Universit\'{e} Paris Diderot, 5 Rue Thomas Mann, F-75013 Paris, France}

\author{S. Villanueva Jr.}
\affiliation{Kavli Institute for Astrophysics and Space Research, M.I.T., Cambridge, MA 02139, USA}

\author{R. A. Street}
\affiliation{LCOGT, 6740 Cortona Drive, Suite 102, Goleta, CA 93117, USA}

\author{Y. Tsapras}
\affiliation{Zentrum f\"ur Astronomie der Universit\"at Heidelberg, Astronomisches Rechen-Institut, M\"onchhofstr. 12-14, D-69120 Heidelberg, Germany}

\author{M. Hundertmark}
\affiliation{Zentrum f\"ur Astronomie der Universit\"at Heidelberg, Astronomisches Rechen-Institut, M\"onchhofstr. 12-14, D-69120 Heidelberg, Germany}

\author{M. Kuzuhara}
\affiliation{Astrobiology Center, National Institutes of Natural Sciences, 2-21-1 Osawa, Mitaka, Tokyo 181-8588, Japan}

\author{M. Fujita}
\affiliation{Department of Earth and Planetary Sciences, Tokyo Institute of Technology, 2-12-1 Ookayama, Meguro-ku, Tokyo 152-8551, Japan}

\author{C. Beichman}
\affiliation{Jet Propulsion Laboratory, California Institute of Technology, Pasadena, CA 91109, USA}
\affiliation{Division of Physics, Mathematics, and Astronomy, California Institute of Technology, Pasadena, CA 91125, USA}
\affiliation{NASA Exoplanet Science Institute, 770 South Wilson Avenue, Pasadena, CA 911225, USA}

\author{J.-P. Beaulieu}
\affiliation{School of Physical Sciences, University of Tasmania, Private Bag 37 Hobart, Tasmania 7001, Australia}
\affiliation{Sorbonne Universites, CNRS, UPMC Univ Paris 06, UMR 7095, Institut d’Astrophysique de Paris, F-75014, Paris, France}

\author{R. Alonso}
\affiliation{Instituto de Astrof\'isica de Canarias, V\'ia L\'actea s/n, E-38205 La Laguna, Tenerife, Spain}
\affiliation{Departamento de Astrof\'isica, Universidad de La Laguna (ULL), E-38206 La Laguna, Tenerife, Spain}

\author{D. E. Reichart}
\affiliation{Department of Physics and Astronomy, University of North Carolina, CB \#3255, Chapel Hill, NC 27599, USA}

\author{N. Kawai}
\affiliation{Department of Physics, Tokyo Institute of Technology, 2-12-1 Ookayama, Meguro-ku, Tokyo 152-8551, Japan}

\author{M. Tamura}
\affiliation{Astrobiology Center, National Institutes of Natural Sciences, 2-21-1 Osawa, Mitaka, Tokyo 181-8588, Japan}
\affiliation{Department of Astronomy, Graduate School of Science, The University of Tokyo, 7-3-1 Hongo, Bunkyo-ku, Tokyo 113-0033, Japan}
\affiliation{National Astronomical Observatory of Japan, 2-21-1 Osawa, Mitaka, Tokyo 181-8588, Japan}

\begin{abstract}
 We report the analysis of additional multiband photometry and spectroscopy and new adaptive optics (AO) imaging of the nearby planetary microlensing event TCP~J05074264+2447555 (Kojima-1), which was discovered toward the Galactic anticenter in 2017 \red{(Nucita et al.)}. We confirm the planetary nature of the light-curve anomaly around the peak while finding no additional planetary feature in this event. We also confirm the presence of apparent blending flux and the absence of significant parallax signal reported in the literature. The AO image reveals no contaminating sources, making it most likely that the blending flux comes from the lens star. The measured multiband lens flux, combined with a constraint from the microlensing model, allows us to narrow down the previously unresolved mass and distance of the lens system. We find that the primary lens is a dwarf on the K/M boundary (0.581~$\pm$~0.033~$M_\odot$) located at $505 \pm 47$~pc, and the companion (Kojima-1Lb) is a Neptune-mass planet (20.0~$\pm$~2.0~$M_\oplus$) with a semi-major axis of 1.08~$^{+0.62}_{-0.18}$ au. This orbit is a few times smaller than those of typical microlensing planets and is comparable to the snow-line location at young ages. We calculate that the a priori detection probability of Kojima-1Lb is only $\sim$35\%, which may imply that Neptunes are common around the snow line, as recently suggested by the transit and radial velocity techniques. The host star is the brightest among the microlensing planetary systems ($K_s=13.7$), offering a great opportunity to spectroscopically characterize this system, even with current facilities.
\end{abstract}


\section{Introduction}

According to core accretion theory, once a protoplanetary core reaches a critical mass of $\sim$\red{10}~$M_\oplus$ by accumulating planetesimals, the protoplanet starts to accrete the surrounding gas in a runaway fashion and quickly becomes a gas giant planet \citep[e.g.,][]{1996Icar..124...62P}. This process can most efficiently happen just outside the snow line, where the surface density of solid materials is enhanced by condensation of ices \citep[e.g.,][]{2004ApJ...604..388I}. Because this process is basically controlled by the mass of the protoplanet, unveiling the planetary mass distribution around the snow line is crucial to understand the planetary formation processes. Recent microlensing surveys have revealed that Neptune-mass-ratio planets are the most abundant in the region several times outside the snow line \citep{2016ApJ...833..145S,2018AcA....68....1U}; however, little is known about the population of low-mass planets just around the snow line.

The microlensing technique is most sensitive to planets with an orbital separation close to the Einstein radius, which is defined by the radius of the ringed image produced when the lens and source stars are perfectly aligned. This size is expressed by 
\footnotesize
\begin{eqnarray}
R_E &=& \sqrt{\frac{4G}{c^2} M_L D_S x(1-x)}\\
&\simeq& 2.9 {\rm au} \left(\frac{M_L}{0.5 M_\odot}\right)^{1/2} \left(\frac{D_S}{8{\rm kpc}}\right)^{1/2} \left[\frac{x(1-x)}{0.25}\right]^{1/2},
\end{eqnarray}
\normalsize
where $M_L$ is the mass of the lens star, $x = D_L/D_S$, and $D_L$ and $D_S$ are the distances to the lens and source stars, respectively.
Assuming that the snow-line distance in a protoplanetary disk can be approximated by $a_{\rm snow} \sim 2.7 {\rm au} \times M_*/M_\odot$, where $M_*$ is the stellar mass \citep{2008ApJ...684..663B}, one can write the ratio of the Einstein radius to the median sky-projected distance of the randomly oriented snow-line orbit, $a_{\rm snow, \perp}  = 0.866 a_{\rm snow}$, as
\footnotesize
\begin{eqnarray}
\frac{R_E}{a_{\rm snow, \perp}} \simeq 2.4 \left(\frac{M_L}{0.5 M_\odot}\right)^{-1/2} \left(\frac{D_S}{8{\rm kpc}}\right)^{1/2} \left[\frac{x(1-x)}{0.25}\right]^{1/2}.
\end{eqnarray}
\normalsize
Thus, the Einstein radius of typical microlensing events toward the Galactic bulge ($M_L\sim$0.5~$M_\odot$, $x \sim 0.5$, and $D_S \sim 8$~kpc), where dedicated microlensing surveys have been conducted, is a few times larger than the snow-line distance \citep[see, e.g., ][for a recent review of microlensing]{2018Geosc...8..365T}.

Because the Einstein radius is scaled by $\sqrt{D_S}$, the planet sensitivity region of microlensing coincides with the location of the snow line \red{when} the distance of the source is \red{an order of magnitude} closer than the distance to the Galactic bulge\red{, i.e., $D_S \sim 1$~kpc}. Although the event rate of such nearby-source microlensing events is \red{expected to be} small \citep[$\sim$23 events yr$^{-1}$;][]{2008ApJ...681..806H}, they can provide a rare opportunity to find and characterize planets just around the snow line. In addition, once such a nearby planetary microlensing event is discovered, it can be an invaluable system that allows spectroscopic follow-up, which is usually difficult for the events observed toward the Galactic bulge.

This is the case for the nearby microlensing event TCP~ J05074264+2447555\footnote{The equatorial and galactic coordinates of this object are ($\alpha$, $\delta$)$_{\rm J2000}$ = (05$^{\rm h}$07$^{\rm m}$42$^{\rm s}$.725, +24$^\circ$47$'$56\farcs37) and ($l$, $b$)$_{\rm J2000}$ = (178$^\circ$.76,  -9$^\circ$.32), respectively.} (hereafter Kojima-1\footnote{Note that \cite{2018MNRAS.476.2962N} nicknamed this event as Feynman-01 in honor of the observatory where the planetary feature was observed. In this paper, we call this event Kojima-1 in honor of Mr. Kojima as the first discoverer of this event. Conventionally, a planetary microlensing event is named after the group(s) that discovers the event itself, rather than the group(s) that detects the planetary feature.}), which was serendipitously discovered during a nova search conducted by an amateur astronomer, Mr. T.~Kojima. On 2017 October 31 UT, he reported an unknown transient event on an $R = 13.6$ mag star toward the Taurus constellation,\footnote{http://www.cbat.eps.harvard.edu/unconf/followups\slash{}J05074264+2447555.html} and later, the microlensing nature of this event was confirmed by photometric and spectroscopic follow-up observations \citep{2017ATel10923....1J,2017ATel10926....1K,2017ATel10919....1M,2017ATel10921....1S}. Moreover, a planetary feature was detected near the peak of the event by the earliest photometric follow-up observations \citep{2017ATel10934....1N}. 

\cite{2018MNRAS.476.2962N} estimated that the distance to the source star is $\sim$700-800~pc. They also fit their own and publicly available light curves with a binary-lens microlens model, finding that the mass ratio of the primary lens to its companion is $(1.1 \pm 0.1) \times 10^{-4}$; i.e., the companion is a planet. However, because of the degeneracy between the absolute mass and distance of the lens system, they estimated them using a stochastic technique based on a Galactic model such that the planetary mass is 9.2$\pm$6.6~$M_\oplus$, the host star's mass is $\sim$0.25~$M_\odot$, and the distance to the system is $\sim$380~pc.
On the other hand, \cite{2019ApJ...871...70D} measured the angular Einstein radius $\theta_{\rm E}$ of this event by observing the separation of the two microlensed source star images using the VLTI/GRAVITY instrument. They confirmed that the \red{$\theta_{\rm E}$ value estimated by \cite{2018MNRAS.476.2962N} is largely consistent with the value measured by VLTI}, although they did not attempt to improve the physical parameters of the lens system using the improved $\theta_{\rm E}$.

Reacting to the discovery of this remarkable event, we started follow-up observations by means of photometric monitoring, high- and low-resolution spectroscopy, and high-resolution imaging to obtain a better understanding of the lens system.

This paper is organized as follows. We describe our follow-up observations and reductions in Section \ref{sec:obs}, and light-curve modeling in Section \ref{sec:lc_modeling}. The properties of the source star and lens system are derived in Sections \ref{sec:source_properties} and \ref{sec:lens_properties}, respectively. We then discuss the possible formation scenario of the planet, detection efficiency of the planet, and capabilities of future follow-up observations of the planetary system in Section \ref{sec:discussions}. We summarize the paper in Section \ref{sec:summary}.

\section{Observations}
\label{sec:obs}

\subsection{Photometric Monitoring}

We conducted photometric monitoring observations of Kojima-1 using 13 ground-based telescopes distributed around the world through the optical ($g$, $r$, $i$, $z_s$, $B$, $V$, $R$, and $I$) and near infrared ($K_s$) bands, as listed in Table \ref{tbl:tel_list}.  The photometric follow-up campaign started on 2017 October 31 and lasted for 76 days until the source's brightness well returned to the original state. The number of observing nights, median observing cadence after removing outliers and time-binning, and median photometric error of each instrument are appended to Table \ref{tbl:tel_list}. 
We note that we triggered the follow-up campaign without knowing the presence of the planetary anomaly, which was first reported on 2017 November 8 \citep{2017ATel10934....1N}. Also, we did not change any observing cadences after the report of the anomaly detection because (1) the anomaly had already finished at the time of the report and therefore no further follow-ups were required for the anomaly itself, and (2) from the beginning, we intended to follow up the event as much as possible until the end of the event, no matter whether a planetary anomaly was detected around the peak or not, to search for new planetary signals. On the other hand, we would have terminated our follow-up campaign by the end of 2017 if the planetary anomaly was not detected, and we extended the campaign for $\sim$2 weeks in reaction to the anomaly detection hoping to place a better constraint on the microlensing light-curve model. We will reflect this point in the calculation of the planet detection efficiency in Section \ref{sec:DE}. We further note that the data from CBABO and SL in the list were also used in \cite{2018MNRAS.476.2962N}; however, we rereduced them with our own photometric pipeline in order to investigate the possible systematics in these data (see below for CBABO and Section \ref{sec:data sets} for SL).

All of the data were corrected for bias and flat-field in a standard manner. To extract the light curves of the event, aperture photometry was performed using a custom pipeline \citep{2011PASJ...63..287F} for the data sets of MuSCAT, MuSCAT2, ISAS, OAOWFC, CBABO, COAST, SL, and MITSuME; {\tt IRAF/APPHOT}\ \footnote{IRAF is distributed by the National Optical Astronomy Observatory, which is operated by the Association of Universities for Research in Astronomy (AURA) under a cooperative agreement with the National Science Foundation.} for Araki; {\tt SExtractor} \citep{1996A&AS..117..393B} for PROMPT-8; {\tt AIJ} \citep{2017AJ....153...77C} for OAR and WCO, and a differential image analysis using the {\tt ISIS} package\ \footnote{http://www2.iap.fr/users/alard/package.html} \citep{1998ApJ...503..325A,2000A&AS..144..363A} was performed for the data set of DEMONEXT. In the case of aperture photometry, comparison stars are carefully selected for each data set depending on the field of view, so that systematics arising from intrinsic variabilities of the comparison stars are minimized.

On the raw images of CBABO obtained on 2017 October 31, the flux counts of the target star were close to the saturation of CCD and were affected by the CCD nonlinearity. We corrected this effect by constructing a pixel-level nonlinearity-correction function using a seventh-order polynomial by minimizing the dispersion of the aperture-integrated light curve of a similar-brightness star in the same field of view (TYC~1849-1592-1).

\begin{figure*}
\begin{center}
\includegraphics[width=18cm]{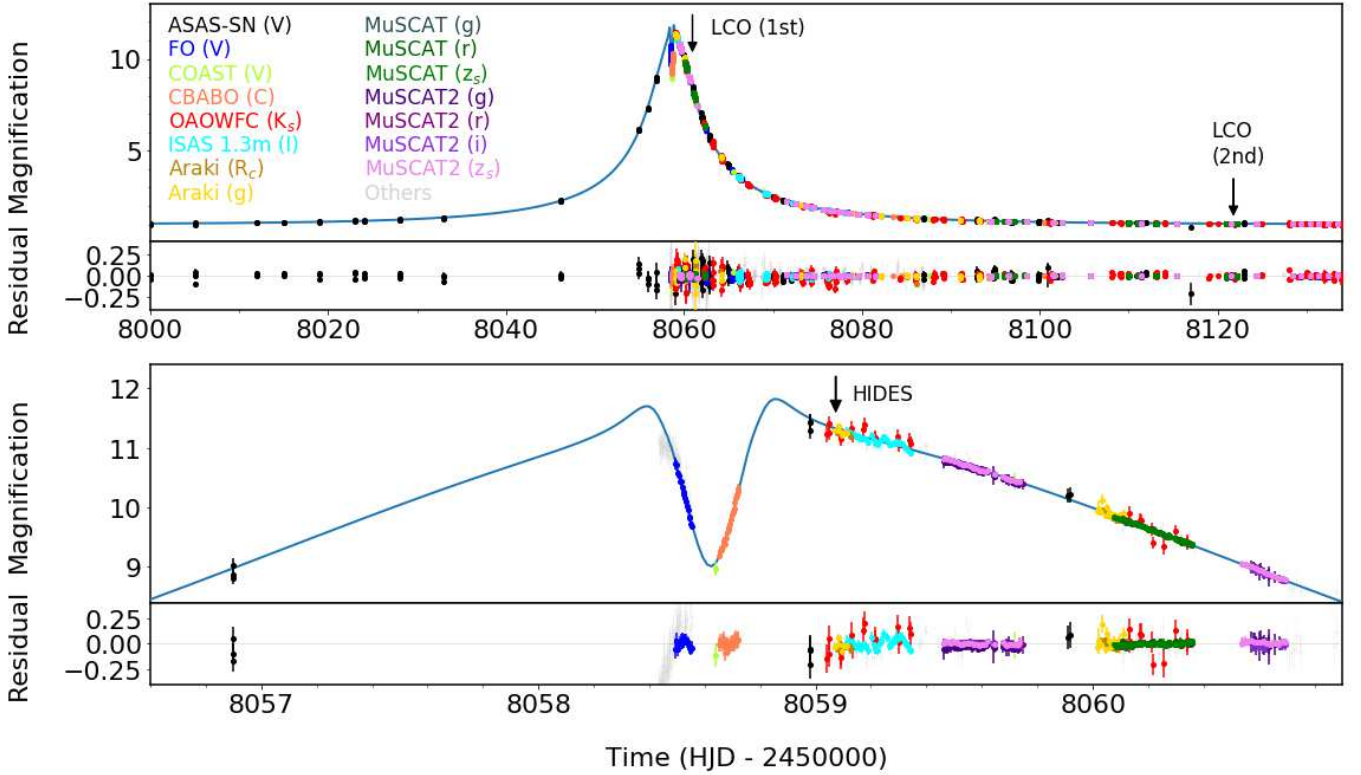}
\caption{(First panel) Light curves of Kojima-1. Colored (including black) and light gray points are the data used for the light-curve fitting and only for the calculation of detection efficiency, respectively. Color legends are shown on the left-hand side. The best-fit microlensing model is indicated by a blue solid line. The times when the two LCO spectra were taken are indicated by arrows.  (Second panel) Residuals from the best-fit model. (Third panel) Zoomed light curves around the peak.  The time when the HIDES spectrum was obtained is indicated by an arrow. (Fourth panel) Residuals for the zoomed light curves.
\label{fig:lc}}
\end{center}
\end{figure*}

The observed light curves are shown in Figure \ref{fig:lc} in magnification scale. While we confirmed the planetary feature around the peak in the data sets of COAST, CBABO, and SL, we did not detect any additional anomaly in the light curves.

\begin{deluxetable*}{lcccccccc}
\tablecaption{List of Photometric Data Sets
\label{tbl:tel_list}}
\tablehead{
\colhead{Abbreviation} & \colhead{Observatory} & \colhead{Telescope}
& \colhead{Field of View} &  \colhead{Filter}  & \colhead{Number of} & \colhead{Number of} & \colhead{Median} & \colhead{Median}\\[-5pt]
(Instrument)\tablenotemark{a}\tablenotemark{b} &&{Diameter}&&&{Nights\tablenotemark{c}} &{Data\tablenotemark{c}}&{Cadence\tablenotemark{c}}&{Flux Error\tablenotemark{c}}\\
 &&(m)&(arcmin$^2$)&&& & (min)& [\%]
}
\tabletypesize{\scriptsize}
\startdata
\multicolumn{2}{l}{{\it Data Sets Obtained or Rereduced in This Work}}\\
{\bf MuSCAT} & NAOJ/Okayama &  1.88  & 6.1 $\times$ 6.1& $g$ &11 & 161 & 10.0 & 0.24\\
&&&&$r$&12 & 163 & 10.0 & 0.16\\
&&&&$z_s$&12 & 196 & 10.0 & 0.30\\
{\bf MuSCAT2} & Teide Observatory & 1.52 & 7.4 $\times$ 7.4 & $g$ & 29 & 331 & 10.0 & 0.38\\
&&&&$r$&27 & 317 & 10.0 & 0.21\\
&&&&$i$&29 & 316 & 10.0 & 0.21\\
&&&&$z_s$&30 & 343 & 10.0 & 0.24\\
{\bf Araki} & Koyama Astronomical Observatory & 1.3 & 12.2 $\times$ 12.2 & $g$ & 12 & 68 & 12.1 & 0.29\\
&&&&$R_c$&12 & 70 & 10.8 & 0.56\\
{\bf ISAS} & JAXA/ISAS & 1.3 & 5.4 $\times$ 5.4 & $I_c$ & 8 & 175 & 10.1 & 0.67\\
{\bf OAOWFC} & NAOJ/Okayama & 0.91  & 28.6 $\times$ 28.6 & $K_s$ & 43 & 202 & 56.0 & 1.95\\
{\bf CBABO} & CBA Belgium Observatory &  0.40 & 12.5 $\times$ 8.4 & Clear & 5 & 30 & 4.9 & 0.77\\
{\bf COAST} & Teide Observatory &  0.35 &33 $\times$ 33 & $V$ & 6 & 7 & --- & 1.18\\
PROMPT-8 & Cerro Tololo Inter-American Observatory  & 0.61 & 22.6 $\times$ 22.6 & $V$ &8 & 64 & 9.7 & 0.83\\
&&&&$R_c$&9 & 79 & 9.7 & 0.69\\
&&&&$I_c$&7 & 70 & 9.7 & 1.21\\
SL & AISAS in Star\'{a} Lesn\'{a} & 0.60 & 14.4 $\times$ 14.4 & $B$ & 3 & 114 & 4.6 & 2.08\\
&&&&$V$ &3 & 198 & 5.2 & 1.18\\
&&&&$R_c$&3 & 121 & 4.8 & 1.63\\
&&&&$I_c$&3 & 177 & 5.2 & 1.37\\
MITSuME & NAOJ/Okayama & 0.50  & 26 $\times$ 26 & $Ic$  &28 & 239 & 13.3 & 1.06\\
DEMONEXT & Winer Observatory & 0.50 & 30.7 $\times$ 30.7 & $I_c$ & 20 & 420 & 10.5 & 2.73\\
OAR & Hankasalmi Observatory & 0.40 & 25 $\times$ 25 &$V$ &4 & 39 & 6.4 & 0.68\\
WCO & Westminster College Observatory & 0.35 & 24 $\times$ 16&CBB &5 & 129 & 9.8 & 0.18\\
\hline
\multicolumn{2}{l}{{\it Public or Published Data Sets}}\\
{\bf FO} & R.P. Feynman Observatory & 0.30 & 27.0 $\times$ 21.6 & $V$ & 5 & 54 & 8.8 & 0.59\\
{\bf ASAS-SN} & Haleakala Observatory & 0.14 & 273 $\times$ 273 & $V$& 44 & 146 & --- & 2.27
\enddata
Notes.
\tablenotetext{a}{The data sets used in the light-curve fitting are shown in bold.}
\tablenotetext{b}{References to the instruments are as follows. MuSCAT: \cite{2015JATIS...1d5001N}; MuSCAT2: \cite{2019JATIS...5a5001N}; OAOWFC: \cite{2016SPIE.9908E..5DY}; MISTuME: \cite{2005NCimC..28..755K},\cite{2010AIPC.1279..466Y}; DEMONEXT: \cite{2018PASP..130a5001V}. }
\tablenotetext{c}{The values for the data after removing outliers and binning time series are reported.}
\end{deluxetable*}

\subsection{High-resolution Spectroscopy}

A high-resolution spectrum was taken in the wavelength range of 4990 -- 7350~\AA\  using the NAOJ 188~cm telescope in Okayama, Japan, and the High Dispersion Echelle Spectrograph \citep[HIDES;][]{2013PASJ...65...15K} on 2017 November 1.6 UT. Two exposures were obtained in the high-efficiency mode (HE mode; $R \sim$ 55,000) with exposure times of 23 and 20 minutes. The data reduction (bias subtraction, flat-fielding, spectrum extraction, and wavelength calibration) was performed by using the IRAF {\tt echelle} package in a standard manner. The signal-to-noise ratio (S/N) ratio of the obtained spectrum is approximately 20--30.

\subsection{Low-resolution Spectroscopy}

Low-resolution spectra ($R \sim$ 500) were taken on 2017 November 3  and 2018 January 3 using the FLOYDS spectrograph mounted on the Las Cumbres Observatory (LCO) 2~m telescope on Haleakala, Hawaii \footnote{More details on the LCO instruments and telescope are available here: https://lco.global/observatory/}. 
The spectral range is about 3200--10000 \AA. Each spectrum was taken with 1000~s exposure with the 1\farcs2 slit. Both spectra were obtained in similar sky conditions, but due to the different magnification at the time of exposure (8.34 and 1.04), both images were obtained with different S/Ns, a range of [50, 250] and [20, 90], respectively.
Both 1D spectra were extracted using the FLOYDS pipeline \footnote{https://github.com/svalenti/FLOYDS\_pipeline}.

\subsection{High-resolution Imaging}
\begin{figure*}
\begin{center}
\includegraphics[width=16cm]{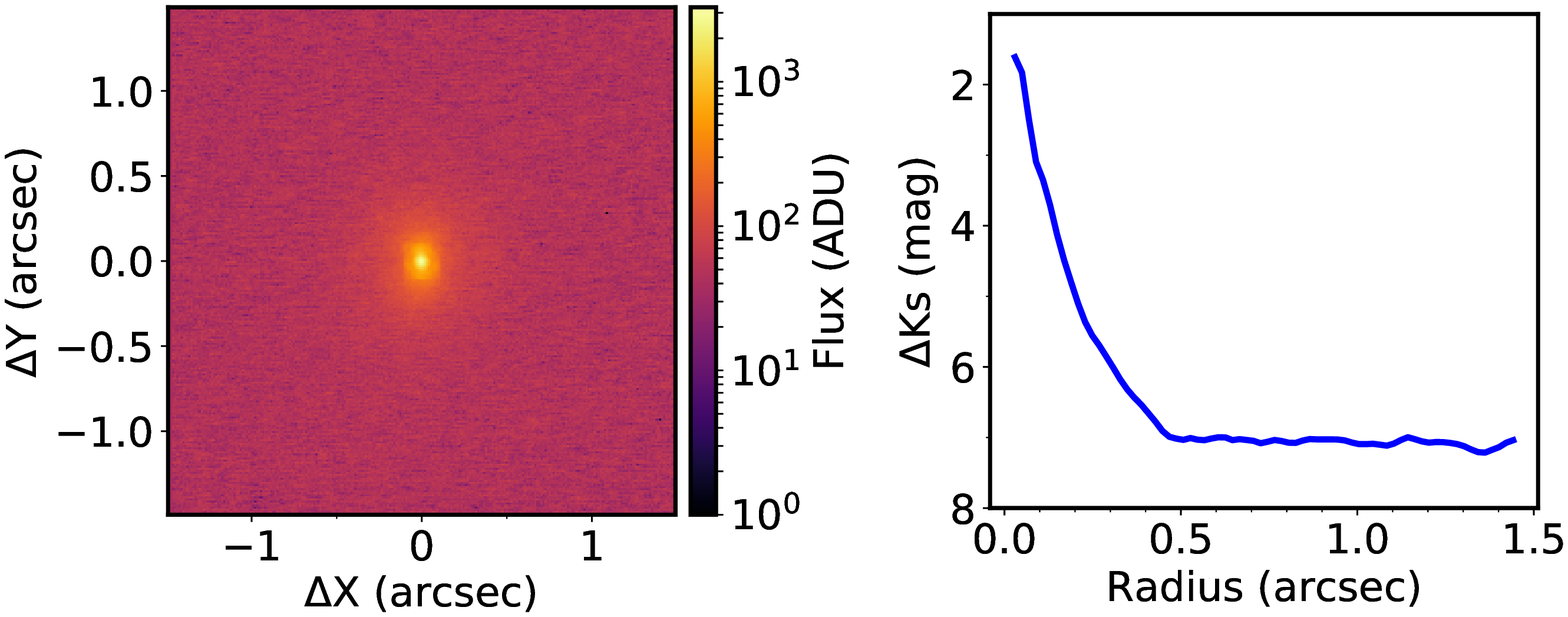}
\caption{(Left) The $K_s$-band AO image of the Kojima-1 object obtained with Keck/NIRC2. (Right) A 5-$\sigma$ contrast curve as a function of the distance from the centroid of the object.
\label{fig:AO}}
\end{center}
\end{figure*}

High resolution images of the event object were obtained using the Keck telescope and NIRC2 instrument on 2018 February 5. Using the narrow camera (pixel scale of 9.94 mas pixel$^{-1}$), 10 dithered images were obtained in the $K_s$ band with the NGS mode, each with an exposure time of 2~s and three co-adds.
The median FWHM of the adaptive optics (AO)-guided stellar point-spread function was 0\farcs06. The raw images were median-combined after bias flat correction, sky subtraction, and stellar position alignment. The combined image and a 5$\sigma$ contrast curve are shown in Figure \ref{fig:AO}. We found no contaminating sources brighter than $K_s = 21$ within the image.

\section{Light-curve Modeling}
\label{sec:lc_modeling}

\subsection{Model Description}
To derive the physical parameters of the lens system, we fit the light curves with a binary-lens microlensing model.
The model calculates the magnification of the source star as a function of time, $A(t)$, which is expressed by the following parameters: the time of the closest approach of the source to the lens centroid, $t_0$; the Einstein radius crossing time, $t_{\rm E}$; the source-lens angular separation at time $t_0$ in units of the angular Einstein radius ($\theta_{\rm E}$), $u_0$; the mass ratio of the binary components, $q$; the sky-projected separation of the binary components in units of $\theta_{\rm E}$, $s$; the angle between the source trajectory and the binary-lens axis, $\alpha$; the angular source radius in units of $\theta_{\rm E}$, $\rho$; and the microlens parallax vector, $\bm{\pi_{\rm E}}$. Here the direction of $\bm{\pi_{\rm E}}$ is the same as the direction of the source's proper motion relative to the lens, and the length of $\bm{\pi_{\rm E}}$, $\pi_{\rm E} \equiv \sqrt{\pi_{\rm E,N}^2 + \pi_{\rm E,E}^2}$,  is equal to the ratio of 1~au to the projected Einstein radius onto the observer plane, where $\pi_{\rm E,N}$ and $\pi_{\rm E,E}$ are the north and east components of $\bm{\pi_{\rm E}}$, respectively. The limb-darkening effect of the source star is modeled by the following formula $I (\theta) = I (0) [1 - u_X (1 - \cos \theta)]$, where $\theta$ is the angle between the normal to the stellar surface and the line of sight, $I (\theta)$ is the stellar intensity as a function of $\theta$, and $u_X$ is a coefficient for filter $X$. 
The observed flux in the $i$th set of instrument and band at time $t$ is expressed by the following linear function $F_i(t) = A(t) \times F_{s,i} + F_{b,i}$, where $F_{s,i}$ and $F_{b,i}$ are the unmagnified source flux and blending flux, respectively, in the $i$th data set.  
Note that the effect of the orbital motion of the planet is not considered in the final analysis because it was not significant in the first trials.

\subsection{Error Normalization}

The initially estimated uncertainties of individual data points are rescaled using the following formula
\begin{eqnarray}
\sigma_i' = k \sqrt{\sigma_i^2 + e_\mathrm{min}^2},
\end{eqnarray}
where $\sigma_i$ is the initial uncertainty of the $i$th data point in magnitude, and $k$ and $e_\mathrm{min}$ are coefficients for each data set. Here the term $e_\mathrm{min}$ represents systematic errors that dominate when the flux is significantly increased. The $k$ and $e_\mathrm{min}$ values are adjusted so that the cumulative $\chi^2$ distribution for the best-fit binary-lens model including the parallax effect sorted by magnitude is close to linear and $\chi^2_{\rm red}$ becomes unity. This process is iterated several times.

In addition, we quadratically add 0.5\% in flux to each flux error for the data points that lie within the anomaly, taking into account the possible intrinsic variability of the target and/or comparison stars. This additional error is important to properly estimate the uncertainties of the model parameters, in particular of $s$, $\rho$, and $\pi_{\rm E}$, which we find are sensitive to this anomaly part and can be biased by even a small systematics of the level of 0.5\% in flux.

\subsection{\red{Data Sets and Fitting Codes}}
\label{sec:data sets}

To save computational time, we restrict the data sets for a light-curve fitting to the ones with relatively high photometric precision with sufficient time coverage and/or have unique coverage in time or wavelength, specifically, the data sets of MuSCAT, MuSCAT2, Araki, ISAS, OAOWFC, CBABO, and COAST. To supplement our data, we also use the $V$-band light curve from All-Sky Automatic Survey for Supernovae \citep[ASAS-SN,][]{2014ApJ...788...48S,2017PASP..129j4502K} (data are extracted from their web site\footnote{https://asas-sn.osu.edu} for the period of 7967 $<$ HJD-2,450,000 $<$ 8123), which covered the entire event with an average cadence of several per night, and the $V$-band light curve capturing the declining part of the anomaly obtained at the R. P. Feynman Observatory (FO) by \cite{2018MNRAS.476.2962N}. 

We note that although the SL data set includes the earliest data points among all of the follow-up observations partly overlapping with the FO data set (HJD-2,450,000 $\sim$ 8058.5), we have not included it in our light-curve modeling for the following reasons. First, when we fit the light curves including this data set, we found that the data points of this data set in the anomaly part have a small systematic trend against the best-fit model. Second, we also found that the $F_s$ and $F_b$ values from this data set, calibrated to standard photometric systems, were discrepant with those from the other same-band data sets at the 2$\sigma$ level,\footnote{Although we found no clear evidence for the cause of this systematics, the stellar positions on the detector moved by $>$50 pixels during the observations, which might cause systematics on the photometry at some level.} even using only the data points that overlap with the FO data set. Because light-curve models are sensitive to the data points in the anomaly part, even a 2$\sigma$ level systematics could cause a tension in the derived parameters.

The light curves are fitted with a binary-microlensing model using a custom code that has been developed for the Microlensing Observations in Astrophysics (MOA) project \citep{2010ApJ...710.1641S}, in which the posterior probability distributions of the parameters are calculated by the Markov chain Monte Carlo (MCMC) method. 
Note that the light curves are also independently analyzed using the pipeline {\tt PyLIMA} \citep{2017AJ....154..203B}, a code developed by \cite{2010ApJ...716.1408B}, and the modeling platform {\tt RTModel}\footnote{http://www.fisica.unisa.it/GravitationAstrophysics\slash{}RTModel.htm} \citep{2018MNRAS.479.5157B} \red{for sanity check}. 

\subsection{\red{Static Model}}

We first fit the light curves with a binary-lens model without the microlens parallax effect (static model), fixing $\pi_{\rm E,E}$ and $\pi_{\rm E,N}$ at zero, to compare with the result of \cite{2018MNRAS.476.2962N}, in which this effect was not taken into account. The median value and 1$\sigma$ confidence interval of the posterior probability distributions of the parameters are listed in Table \ref{tbl:lc_fit}. We recover the two degenerate models found by \cite{2018MNRAS.476.2962N} (models $a$ and $b$), in which only $s$ is slightly different and all the other parameters are almost identical between the two models. The best-fit $\chi^2$ values are almost the same between the two models, namely, 2557.5 and 2557.4 for models $a$ and $b$, respectively, for the degrees of freedom (dof) of 2578.  In Table \ref{tbl:lc_fit}, we report the values derived only for model $b$ for all parameters except for $s$, and hereafter, we will discuss them along with this model unless otherwise described.

Our derived values are consistent with those of \cite{2018MNRAS.476.2962N} within 2$\sigma$ for all parameters except for $u_0$, $s$, and $\rho$, for which the discrepancy can be attributed to the following differences between our and their data sets: (1) we correct the detector's nonlinearity effect in the CBABO data set, (2) we omit the SL data set from our modeling due to apparent systematics, and (3) we have a larger number of data points with a longer baseline.

\subsection{\red{Parallax Model}}
\label{sec:parallax}

\subsubsection{\red{Without Informative Prior}}
\label{sec:parallax_without_prior}

To search for a signal of the parallax effect, we fit the light curves letting $\pi_{\rm E,E}$ and $\pi_{\rm E,N}$ be free, first without any informative priors. The derived values and uncertainties are reported in Table \ref{tbl:lc_fit}. 
From this fit, we marginally detect a nonzero $\pi_{\rm E}$ value of $0.34\ ^{+0.34}_{-0.20}$.
However, the $\chi^2$ improvement of the best-fit parallax model over the static model is 14.4, which is not significant enough to claim a detection of the parallax signal, given that the Bayesian information criterion (BIC $\equiv \chi^2 + k \ln N_{\rm data}$, where $k$ is the number of free parameters and $N_{\rm data}=2615$ is the number of data points) for the parallax model is larger (worse) than the static model by 1.3.

\begin{figure}
 \begin{center}
 \includegraphics[width=8cm]{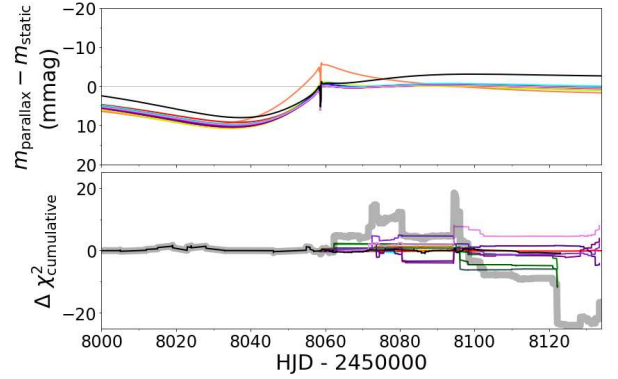}
 \caption{\red{(Top) Difference of best-fit model magnitudes between the parallax and static models for individual data sets, where the color codes are the same as in Figure \ref{fig:lc}. (Bottom) Difference of cumulative $\chi^2$ between the parallax and static models for individual data sets (thin colored lines) and all data sets (thick gray line), where negative means that the parallax model is preferred. The color codes are the same as in Figure \ref{fig:lc}.}
  \label{fig:para_vs_static}}
 \end{center}
 \end{figure}

We also check where the marginal parallax signal comes from. In the top panel of Figure \ref{fig:para_vs_static}, we show the magnitude differences between the best-fit static and parallax models for individual data sets, which indicate that the largest difference arises around $\sim$20 days before the peak, yet the difference is at most at the $\sim$10~mmag level. On the other hand, in the bottom panel of Figure \ref{fig:para_vs_static}, we show the difference of cumulative $\chi^2$ between the two models as a function of time.
This plot indicates that the most of the $\chi^2$ improvements comes from 
only two epochs of the MuSCAT data (from three different bands), where the model magnitudes differ by only $\sim$1 mmag. Thus, the likely origin of the parallax signal is due to systematics in the data at these two epochs, which might arise from the instrument, variability of atmospheric transparency, and/or stellar activity. Therefore, the observed marginal signal of the parallax effect should be treated with caution. Nevertheless, the data still allow us to place an upper limit on $\pi_{\rm E}$ (Section \ref{sec:pi_E_upper_limit}) and constrain the direction of $\bm{\pi_{\rm E}}$ (Section \ref{sec:parallax_with_thetaE_Phipi}).

The result that a significant parallax signal is absent is consistent with the result of \cite{2019ApJ...871...70D}, who also did not detect a significant parallax signal from a single-lens model fit (for the ``luminous-lens'' case in their paper). \cite{2019ApJ...871...70D} described the reasons why the parallax signal in this event is not obvious, which are summarized as follows: (1) the event is quite short compared to a year, (2) it lies quite close to the ecliptic plane, (3) it peaked only 5 weeks \footnote{\cite{2019ApJ...871...70D} erroneously stated it to be 3 weeks. } before opposition, and (4) the lens-source relative proper motion points roughly south. The combination of these factors weakens the parallax signal in the light curve by a factor of $\sim$10 compared to the most favorable case \citep{2019ApJ...871...70D}.

\subsubsection{\red{With Informative Prior on $\theta_{\rm E}$}}

From the light-curve fitting with the parallax model, $\rho$ is measured to be $3.2\ ^{+0.9}_{-1.3} \times 10^{-3}$. This $\rho$ value allows the derivation of the angular Einstein radius $\theta_{\rm E}$ via the relation of $\theta_{\rm E} \equiv \theta_*/\rho$, where $\theta_*$ is the angular radius of the source star. The $\theta_*$ value is estimated to be $8.65 \pm 0.06$~$\mu$as using the procedure described in Section \ref{sec:Ds_and_theta_s}, which leads to $\theta_{\rm E} = \redred{2.7^{+1.9}_{-0.6}}$ mas. On the other hand, the $\theta_{\rm E}$ of the same event was independently and much more precisely determined to be $1.883 \pm 0.014$~mas (in the case of a luminous lens) by \cite{2019ApJ...871...70D} by spatially resolving the two microlensed images during the event. This information can be used to further constrain $\rho$ and some other parameters that are correlated with $\rho$ (in particular, $s$).

Using $\theta_{\rm E}=1.883 \pm 0.014$~mas (in the form of $\rho = \theta_*/\theta_{\rm E}$) as an informative prior, we iteratively fit the light curves refining $\theta_*$ through the process described in Section \ref{sec:Ds_and_theta_s}. The improved parameter values are appended to Table \ref{tbl:lc_fit}, in which notable improvements can be seen in $\rho$, $s$, and $\theta_{\rm E}$. On the other hand, the $\theta_{\rm E}$ prior has not changed the significance of the parallax signal.
 
\subsubsection{\red{Upper Limit on $\pi_{\rm E}$}}
\label{sec:pi_E_upper_limit}
  
From the VLTI observation, \cite{2019ApJ...871...70D} also constrained the direction of $\bm{\pi_{\rm E}}$ ($\Phi_\pi$) into two directions, $193\fdg5 \pm 0\fdg4$ and $156\fdg7 \pm 0\fdg4$ from north to east (for the luminous-lens model). To put an upper limit on $\pi_{\rm E}$ utilizing the prior information of $\Phi_\pi$, we draw $\chi^2$ maps on a grid of $\pi_{\rm E,E}$ and $\pi_{\rm E,N}$. We grid $\pi_{\rm E,E}$ and $\pi_{\rm E,N}$ by a grid size of 0.1 in the ranges of $-0.7 \leqq \pi_{\rm E,E} < 0.7$ and $-1.5 \leqq \pi_{\rm E,N} < 1.5$, and fit the light curves using the $\theta_{\rm E}$ prior while fixing $\pi_{\rm E,E}$ and $\pi_{\rm E,N}$ at each grid-point value. In the left panel of Figure \ref{fig:piEE_vs_piEN}, we show $\Delta \chi^2$ maps on the $\pi_{\rm E,E}$--$\pi_{\rm E,N}$ plane calculated from all data sets, where $\Delta \chi^2$ is the difference of $\chi^2$ between each grid point and ($\pi_{\rm E,E}$, $\pi_{\rm E,N}$) = (0, 0). The minimum-$\chi^2$ (darkest red) region is not coincident with the two solutions of $\Phi_\pi$ (indicated by cyan lines), probably due to the systematics in the light curves discussed before. Note that the reason why the negative $\Delta \chi^2$ region is elongated almost along the $\pi_{\rm E,N}$ direction (only $\pi_{\rm E,E}$ is well constrained) is that the direction of Earth's acceleration is almost parallel to the direction of $\pi_{\rm E,E}$ \footnote{The ecliptic coordinate of the event is ($\beta$, $\lambda$) = (78$^\circ$, 1$\fdg$9), which is close to (90$^\circ$, 0$^\circ$) where the direction of Earth's acceleration is parallel to east-west.}.
On the other hand, the right panel of the same figure shows a $\Delta \chi^2$ map that is calculated only using the $\chi^2$ values from the ASAS-SN data set, which covers the region where the parallax signal is maximized and is thus robust for a parallax signal against the systematics. In this map, although the minimum-$\chi^2$ region is not localized, the intersection between the $\Phi_\pi$ solutions and some $\Delta \chi^2$ contour can still be used to put an upper limit on $\pi_{\rm E}$. The contour of $\Delta \chi^2 = 9$ (white) intersects with the $\Phi_\pi \sim 156\fdg7$ and $\Phi_\pi \sim 193\fdg5$ lines (cyan) at the grid points that correspond to $\pi_{\rm E}$=1.1 and 0.5, respectively. We conservatively adopt 1.1 as a 3$\sigma$ upper limit on $\pi_{\rm E}$.

 \subsubsection{\red{On the Direction of $\bm{\pi_{\rm E}}$}}
 \label{sec:parallax_with_thetaE_Phipi}

\begin{figure*}
 \begin{center}
\includegraphics[width=14cm]{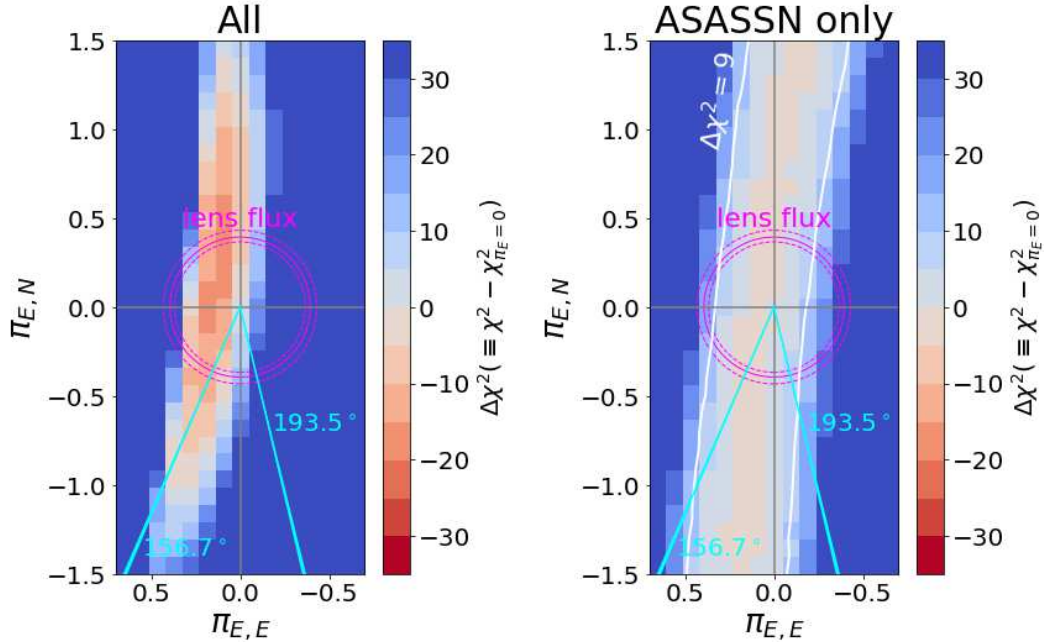}
 \caption{\red{(Left) The $\Delta \chi^2$ map for $\pi_{\rm E,E}$ and $\pi_{\rm E,E}$, where $\Delta \chi^2$ is the $\chi^2$ difference between each grid point and ($\pi_{\rm E,E}$, $\pi_{\rm E,N}$) = (0, 0), calculated using all data sets. The two $\Phi_\pi$ solutions derived from the VLTI observation by \cite{2019ApJ...871...70D} are indicated by cyan lines. The magenta solid and dotted circles correspond to the contours of $\pi_{\rm E} = 0.39\ ^{+0.04}_{-0.03}$, which are expected from the lens flux (see text for details). (Right) Same as the left panel but calculated only using the $\chi^2$ of the ASAS-SN data set. The white solid lines are the contour for $\Delta \chi^2$=9. We estimate the 3$\sigma$ upper limit of $\pi_{\rm E}$ to be 1.1 from the intersection between the white and cyan lines.}
 \label{fig:piEE_vs_piEN}}
 \end{center}
 \end{figure*}

As will be discussed in Section \ref{sec:from_lens_flux}, under the condition of $\pi_{\rm E} < 1.1$, it is most likely that the blending flux detected in the light curves comes from the lens star independently on the $\Phi_\pi$ value, and this lens flux allows us to derive the mass of the lens star to be $M_L = 0.590\ ^{+0.042}_{-0.051} M_\odot$. This lens mass, combined with $\theta_{\rm E}$, predicts the $\pi_{\rm E}$ value using the following relation
\begin{eqnarray}
\label{eq:pi_E}  \pi_{\rm E} =  \frac{\theta_{\rm E}}{\kappa M_L},
\end{eqnarray}
\red{where $\kappa \equiv 4G/c^2$, $G$ is the gravitational constant, and $c$ is the speed of light. This gives $\pi_{\rm E} = 0.39\ ^{+0.04}_{-0.03}$, which is indicated by magenta solid (median) and dotted (1$\sigma$ boundary) contours in  Figure \ref{fig:piEE_vs_piEN}.
In the $\Delta \chi^2$ map for all data sets (left panel of Figure \ref{fig:piEE_vs_piEN}), the $\Delta \chi^2$ value at the grid point that satisfies both $\pi_{\rm E} \sim 0.39$ and $\Phi_\pi \sim 156\fdg7$ is $-16$, which is smaller than the counterpart that satisfies both $\pi_{\rm E} \sim 0.39$ and $\Phi_\pi \sim 193\fdg5$ by 40. This $\chi^2$ difference nominally rules out the $\Phi_\pi=193\fdg5$ solution.}

This outcome, however, could be affected by systematics in the light curves. To test this possibility, we also check the $\Delta \chi^2$ map calculated only using the $\chi^2$ values from the ASAS-SN data set (right panel of Figure \ref{fig:piEE_vs_piEN}). We find that the $\Phi_\pi=156\fdg7$ solution is preferred over the other solution with a $\chi^2$ improvement of $\sim$5, which, although marginal, supports the outcome obtained from all data sets.

Considering the above evidence, we adopt the $\Phi_\pi=156\fdg7$ solution for further analysis. To derive the final posteriors of the parameters, taking into account correlations between the parallax parameters ($\pi_{\rm E,E}$ and $\pi_{\rm E,N}$) and others and using all of the informative prior information, we rerun the MCMC analysis, letting $\pi_{\rm E,E}$ and $\pi_{\rm E,N}$ be free and imposing priors on $\theta_{\rm E}$ and $\Phi_\pi$ with Gaussian distributions of $\theta_{\rm E} = 1.883 \pm 0.014$~mas and $\Phi_\pi = 156\fdg7 \pm 0\fdg4$. The results are reported in Table \ref{tbl:lc_fit}. We note that if the other solution of $\Phi_\pi$ is adopted, then the light-curve fit gives slightly larger values of the blending flux, leading to an $\sim$10\% increase of $M_L$. This, however, does not change the conclusion of this paper much. This $\Phi_\pi$ value can be confirmed in the future by directly measuring the lens-source relative position from high-spatial resolution images.
 
\begin{center}
\begin{deluxetable*}{ccccccc}
\tablecaption{Best-fit Parameter Values of Binary-lens Microlensing Models. 
\label{tbl:lc_fit}}
\tablehead{
\colhead{Parameter \tablenotemark{a}} & \colhead{Unit} & \colhead{\cite{2018MNRAS.476.2962N}}  & \colhead{\red{Static}} & \colhead{Parallax} & \colhead{Parallax} & \colhead{\red{Parallax}} \\
& & & & & \red{w/ $\theta_{\rm E}$ Prior} & w/ $\theta_{\rm E}$, $\Phi_\pi$ Priors
}
\tabletypesize{\scriptsize}
\startdata
$t_0$ & HJD & $0.75 \pm 0.01$ & \red{$0.7353\ \pm 0.0076$} & $0.7395\ \pm 0.0073$ & \red{$0.7396\ \pm 0.0073$} & $0.7403\ \pm  0.0074$ \\
& $-$2,458,058 &&&&&\\
$t_{\rm E}$ & days & $26.4 \pm 0.9$ & \red{$27.44\ \pm 0.07$} & $27.19\ \pm 0.15$ & \red{$27.18\ \pm 0.14$} & $27.25 \pm 0.09$ \\
$u_0$ & 10$^{-2}$ & $\red{9.3} \pm 0.1$ \tablenotemark{b} & \red{$8.858\ ^{+0.031}_{-0.034}$} & $8.925\ \pm 0.043$ & \red{$8.927\ \pm 0.042$} & $8.935\ \pm 0.038$\\
$q$ & 10$^{-4}$& $1.1 \pm 0.1$ & \red{$1.058\ ^{+0.068}_{-0.074}$} & $1.075\ ^{+0.066}_{-0.073}$ & \red{$1.031\ ^{+0.078}_{-0.084}$} & $1.027\ ^{+0.078}_{-0.084}$ \\
$s$ \red{(model $a$)} & & $0.935 \pm 0.004$ & \red{$0.9207\ ^{+0.0045}_{-0.0040}$} & \red{$0.9204\ ^{+0.0040}_{-0.0038}$} & \red{$0.9263\ \pm 0.0018$} & $\red{0.9264\ \pm 0.0018}$\\
$s$ \red{(model $b$)} & & $0.975 \pm 0.004$ & \red{$0.9944\ ^{+0.0041}_{-0.0046}$} & $0.9941\ ^{+0.0042}_{-0.0045}$ & \red{$0.9874\ \pm 0.0018$} & $0.9873\ \pm 0.0018$ \\
$\alpha$ & rad & $\red{4.767} \pm 0.007$ \tablenotemark{b} & \red{$4.7594\ \pm 0.0030$} & $4.7610\ \pm 0.0030$ & \red{$4.7604\ \pm 0.0028$} & $4.7604 \pm 0.0028$ \\
$\rho$ & $10^{-3}$ & $6.0 \pm 0.8$ & \red{$3.2\ ^{+0.9}_{-1.3}$} & $3.2\ ^{+0.9}_{-1.3}$ & \red{$4.568\ \pm 0.070$} & $4.567 \pm 0.071$ \\
$\pi_{\rm E,E}$ && -- & -- & $0.071\ ^{+0.072}_{-0.064}$ & \red{$0.0693\ ^{+0.070}_{-0.063}$} & $0.143\ ^{+0.061}_{-0.053}$\\
$\pi_{\rm E,N}$ && -- & -- & $0.17\ \pm 0.45$ & \red{$0.19\ \pm 0.45$} & $-0.33\ ^{+0.12}_{-0.14}$ \\
\red{$\chi^2_{\rm min}$ / dof} & & -- & \red{2557.4 / 2578} & \red{2543.0 / 2576}& \red{2546.5 / 2577} & \red{2550.7 / 2578}\\
\hline
$\pi_{\rm E}$  && -- & -- & $0.34\ ^{+0.34}_{-0.20}$ \redred{\tablenotemark{c}} & $0.35\ ^{+0.34}_{-0.20}$ \redred{\tablenotemark{c}} & $0.36\ ^{+0.16}_{-0.13}$ \\
$\theta_*$ & $\mu$as  & -- & \red{$8.59 \pm 0.06$} & $8.65 \pm 0.06$ & \red{$8.63 \pm 0.06$} & $8.63 \pm 0.06$\\
$\theta_{\rm E}$ & mas & $1.45 \pm 0.25$ & \red{$2.63\ ^{+1.77}_{-0.58}$} & \redred{$2.68^{+1.87}_{-0.59}$}  & \red{$1.890 \pm 0.032$} & $1.890 \pm 0.032$\\
\enddata
Notes.
\tablenotetext{a}{\red{The values for the two models (model $a$ and $b$) are basically identical except for $s$, for which both values are presented. Only the values for model $b$ are presented for the other parameters. }}
\tablenotetext{b}{\red{For ease of comparison, we multiply the $u_0$ and increment $\alpha$ reported in the literature by $-1$ and $\pi$, respectively. The geometry is identical to this transformation.}}
\tablenotetext{c}{\redred{Because $\pi_{\rm E,E}$ and $\pi_{\rm E,N}$ take both positive and negative values, the median value of $\pi_{\rm E}$ does not coincide with $\sqrt{ \left<\pi_{\rm E,E}\right>^2 + \left<\pi_{\rm E,N}\right>^2}$, where $\left<\pi_{\rm E,E}\right>$ and $\left<\pi_{\rm E,N}\right>$ are the median values of $\pi_{\rm E,E}$ and $\pi_{\rm E,N}$, respectively.}}
\end{deluxetable*}
\end{center}

\begin{figure}
\begin{center}
\vspace{20pt}
\includegraphics[width=8cm]{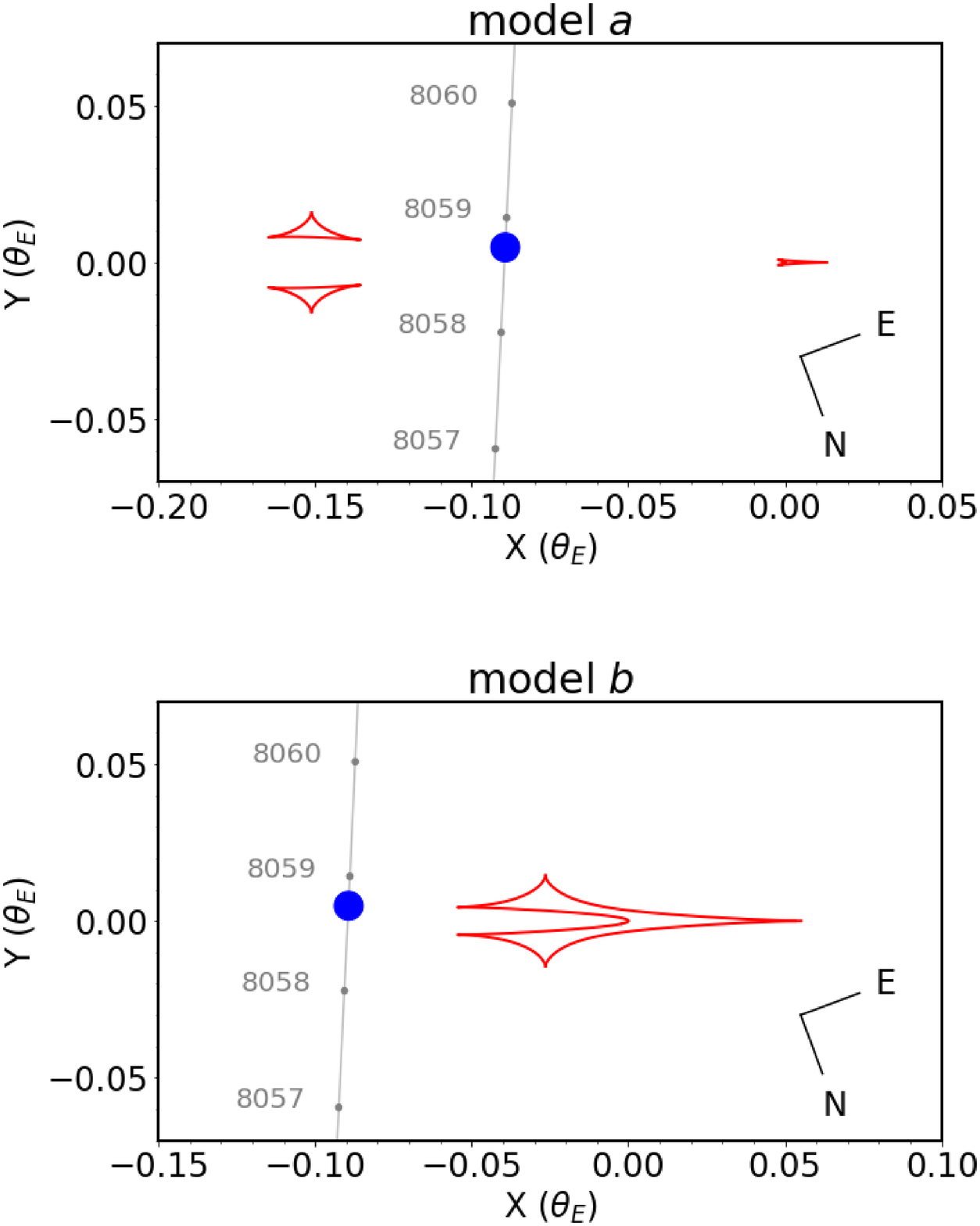}
\caption{Caustic (red) and source trajectory (gray) of the \red{two degenerated} microlensing model\red{s $a$ (top) and $b$ (bottom)}. The time ticks are given by small gray circles. The blue circle represents the source size and position at time $t=t_0$.
\label{fig:caustic}}
\end{center}
\end{figure}

\section{Properties of the Source Star}
\label{sec:source_properties}

In this section, we will derive the properties of the source star, in particular, the source's angular radius $\theta_*$ and the distance to the source star $D_S$, the former of which is tied to $\theta_{\rm E}$ by the relation of $\theta_{\rm E} = \theta_*/\rho$. We measure these values from the brightness of the source star derived from the light-curve fitting with the aid of the spectroscopic information and the extinction from the {\it Gaia} Dada Release 2 (DR2).

\subsection{High-resolution Spectrum} 

The spectroscopic properties of the source star are initially estimated from the HIDES spectrum in the wavelength region of 5000--5900 \AA. Note that the spectrum in longer wavelengths is not used to avoid a significant fringe effect.
Because the spectrum was taken at a time when the source was magnified by a factor of 10, the flux contamination from other objects into the source's spectrum is negligibly small, with a fraction of less than 0.4\% in this wavelength range. We also note that the spectrum does not show any sign of a companion star, i.e., a split of lines due to differential radial velocity. 
Using the spectral fitting tool {\sc SpecMatch-Emp} \citep{2017ApJ...836...77Y}, which matches an observed spectrum with empirical spectral libraries, we estimate the stellar effective temperature, radius, and metallicity to be $T_\mathrm{eff} = 6303 \pm 110$~K, $R_S = 1.56 \pm 0.25$~$R_\odot$, and [Fe/H]=$-0.11 \pm 0.08$, respectively.
This result indicates that the source star is a main-sequence late-F dwarf.

\subsection{Low-resolution Spectrum}

\begin{figure}[h]
\includegraphics[width=9cm]{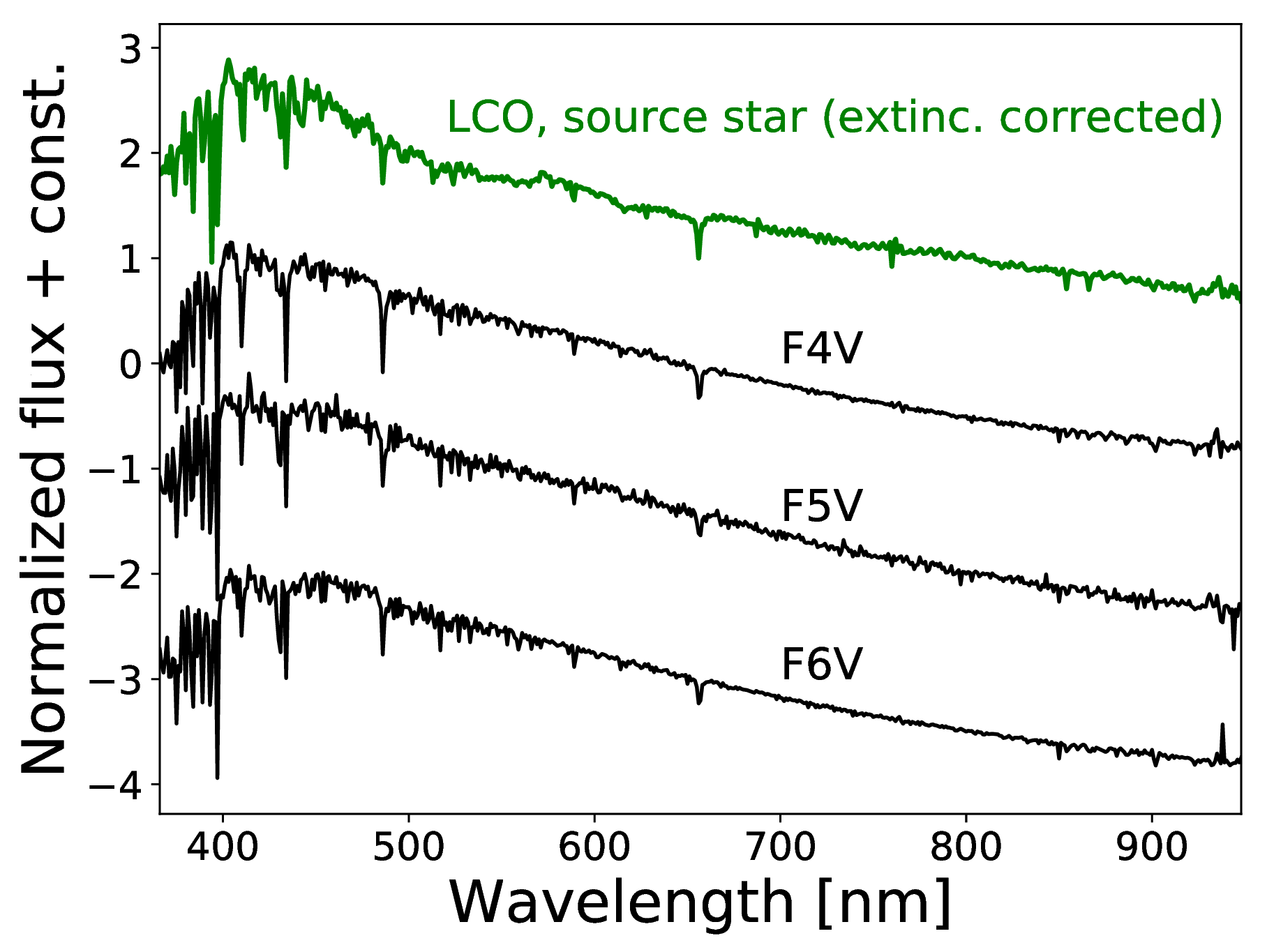}
\caption{Low-resolution spectrum of the source star extracted and extinction-corrected from the LCO spectra (green), along with empirical spectral templates of F4V, F5V, and F6V stars from \cite{2017ApJS..230...16K} (black, top to bottom).
\label{fig:LCO_spec}}
\end{figure}

The two LCO spectra were taken at the magnifications of $A_1=8.34$ and $A_2=1.04$, with which the flux contamination from the lens star, in particular for the wavelength of $\gtrsim$700~nm, is not negligible.
Nevertheless, we can extract the source spectrum from the observed spectra using the equation
$f_{s, \lambda} = (f_{1,\lambda} - f_{2,\lambda})/(A_1-A_2)$, where $f_{1,\lambda}$ and $f_{2,\lambda}$ are the fluxes at the wavelength $\lambda$ in the first- and second-epoch spectra, respectively.
We correct the interstellar extinction in the source spectrum and compare it with empirical spectral templates of \cite{2017ApJS..230...16K}, as shown in Figure \ref{fig:LCO_spec}, finding that the source's spectral type is F5V $\pm$ 1 subtype. This result is consistent with that obtained from the HIDES spectrum.

\subsection{Extinction Estimated from the {\it Gaia} DR2}

The interstellar extinction toward the source star is initially estimated using the {\it Gaia} DR2 \citep{2016A&A...595A...1G,2018A&A...616A...1G}, in which the trigonometric parallax ($\pi$) and extinction in the {\it Gaia} band ($A_G$) are both recorded for a subset of relatively bright and nearby stars. Although the uncertainties of individual $A_G$ values are large, an ensemble of $A_G$ can be used to estimate the averaged $A_G$ value in the field because the uncertainties are dominated by statistical errors \citep{2018A&A...616A...1G}. 

\begin{figure}[h]
\begin{center}
\includegraphics[width=8cm]{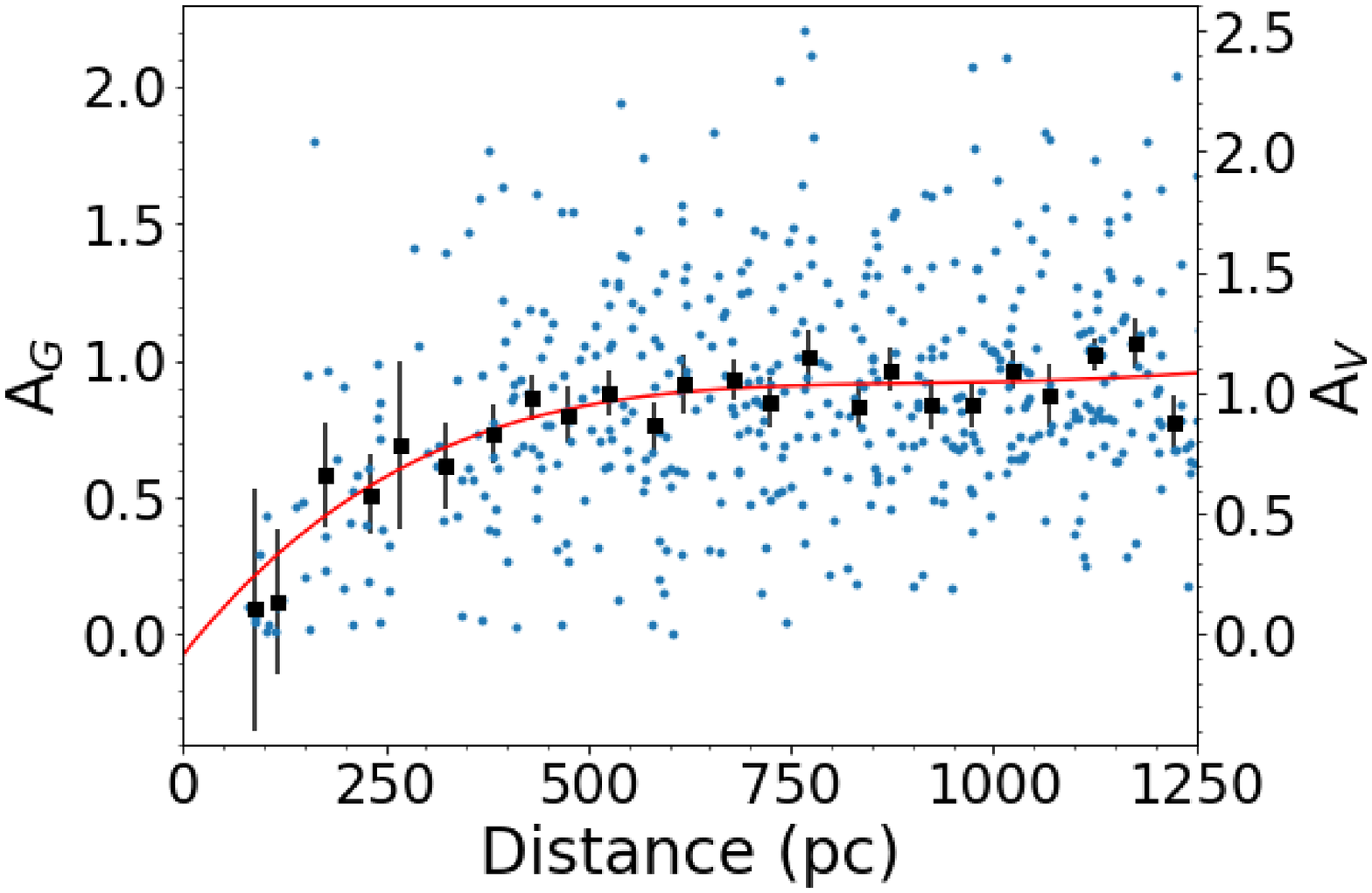}
\caption{
Extinction in the {\it Gaia} (left-hand axis; $A_G$) or visible (right-hand axis; $A_V$) band as a function of distance for stars within 30$'$ in radius from the source position extracted from the {\it Gaia} DR2.
Blue dots are the data for individual stars, and black squares are the binned values with a bin size of 50 pc, where the error bars represent the standard deviation divided by the square root of the number of data points. The red curve indicates the best-fit, fourth-order polynomial function.
\label{fig:AG}}
\end{center}
\end{figure}

First, from the {\it Gaia} DR2, we extract stars that lie within 30$'$ of the source position, have records of both $\pi$ and $A_G$, and have $\pi > 0.5$~mas with a fractional uncertainty of less than 20\%. Next, all of the data are divided by distance into bins with a width of 50~pc. The mean and 1$\sigma$ error (standard deviation divided by the square root of the number of data points) for each bin are calculated, where the median 1$\sigma$ error is $\sim$0.10. The binned data are then fitted with a fourth-order polynomial function of the distance, which gives
\begin{eqnarray}
A_G &=& -7.4918 \times 10^{-2} +  3.6988 \times 10^{-3} D \nonumber \\
&& -5.1142 \times 10^{-6} D^2 + 3.0569 \times 10^{-9} D^3 \nonumber \\
&& -6.4472 \times 10^{-13} D^4,
\label{eq:Av}
\end{eqnarray}
where $D$ is the distance from the Earth. We plot the individual and binned $A_G$ data along with the derived function in Figure \ref{fig:AG}. 
We also calculate the ratio of $A_G$ to $A_V$, which is  the extinction in the $V$ band, to be 1.13, assuming the extinction law of \cite{1989ApJ...345..245C} with $R_V \equiv A_V/E(B-V) = 3.1$.

\subsection{Distance and Angular Radius}
\label{sec:Ds_and_theta_s}

Although the trigonometric parallax of an object at the same coordinates as Kojima-1 was measured by {\it Gaia} to be $1.45 \pm 0.03$~mas, this value does not represent the true trigonometric parallax of the source star but is biased by the foreground lens star. Based on the multiband measurements of $F_s$ and $F_b$, we estimate that the flux ratio of the lens to the source stars in the {\it Gaia} band is $\sim$5\%, assuming that $F_b$ comes entirely from the lens star (see Section \ref{sec:contami}).  On the other hand, the {\it Gaia} DR2 data were acquired during the period between 3.3 and 1.4 yr before the peak of the event, which translates to lens-source separations of $\sim$83 and $\sim$35~mas, respectively. Because the image resolution of {\it Gaia} is 250 mas $\times$ 85 mas, this lens flux fully contaminated to the {\it Gaia} images, substantially changing its position relative to the source star. Therefore, it is not possible to estimate the effect of the lens-flux contamination on the measured parallax without knowing the respective times of the time series of {\it Gaia} astrometric data.

We instead estimate the distance ($D_S$) and angular radius ($\theta_*$) of the source star using the spectral energy distribution (SED) as follows.
First, we calibrate the source fluxes, $F_s$, in the $g$, $r$, $i$, and $z_s$ bands of MuSCAT and MuSCAT2 to the SDSS $g'$, $r'$, $i'$, and $z'$ magnitudes, respectively. We also convert the $F_s$ in the $V$ band of ASAS-SN to the Johnson $V$ magnitude and calibrate the $F_s$ in the $K_s$ band of OAOWFC to the 2MASS $K_s$ magnitude (Table \ref{tbl:source}). The calibrated magnitudes are then converted into flux densities to create the SED.
Next, we fit the SED with the synthetic spectra of BT-Settl \citep{2012RSPTA.370.2765A} using the following parameters: the stellar effective temperature $T_\mathrm{eff}$, radius $R_S$, metallicity [M/H], $A_V$ to the source star $A_{V,S}$, and $D_S$. For a given set of $R_S$ and [M/H], log surface gravity ($\log g$) is calculated using an empirical relation of \cite{2010A&ARv..18...67T}, and from a set of $T_\mathrm{eff}$, [M/H], and $\log g$, a synthetic spectrum is created by linearly interpolating the grid models. The synthetic spectrum is then scaled by $(R_S/D_S)^2$ and reddened using a given $A_{V,S}$ value and $R_V = 3.1$ to fit the observed SED.
We perform MCMC to calculate the posterior probability distribution of each parameter using the {\tt emcee} code \citep{2013PASP..125..306F}. 
In the MCMC sampling, Gaussian priors are applied to the parameters $T_\mathrm{eff}$, $R_S$, [M/H], and $A_{V,S}$ by adding penalties to the $\chi^2$ value as 
\begin{eqnarray}
\chi^2 &=& \sum_{\lambda} \frac{(f_\mathrm{obs, \lambda} - f_\mathrm{model, \lambda})^2}{\sigma_{f_\mathrm{obs, \lambda}}^2} \nonumber\\
&& + \sum_{i} \frac{(X_i - X_{i, \mathrm{prior}})^2}{\sigma_{X_{i, \mathrm{prior}}}^2},
\end{eqnarray}
where $f_\mathrm{obs, \lambda}$, $\sigma_{f_\mathrm{obs, \lambda}}$, and $f_\mathrm{model, \lambda}$ are the observed flux density, its 1$\sigma$ uncertainty, and the model flux density, respectively, for a band $\lambda$, and $X_i$ denotes one of the parameters among $T_\mathrm{eff}$, $R_S$, [M/H], and $A_V$.
For the priors of $T_\mathrm{eff}$, $R_S$, and [M/H], the values derived from the HIDES spectrum are used, where [M/H] and [Fe/H] are assumed to be identical.
As for $A_{V,S}$, the prior value is evaluated using Equation (\ref{eq:Av}) for a given $D_S$, and 0.10 is taken as the 1$\sigma$ uncertainty.

\begin{figure*}[h]
\begin{center}
\includegraphics[width=12cm]{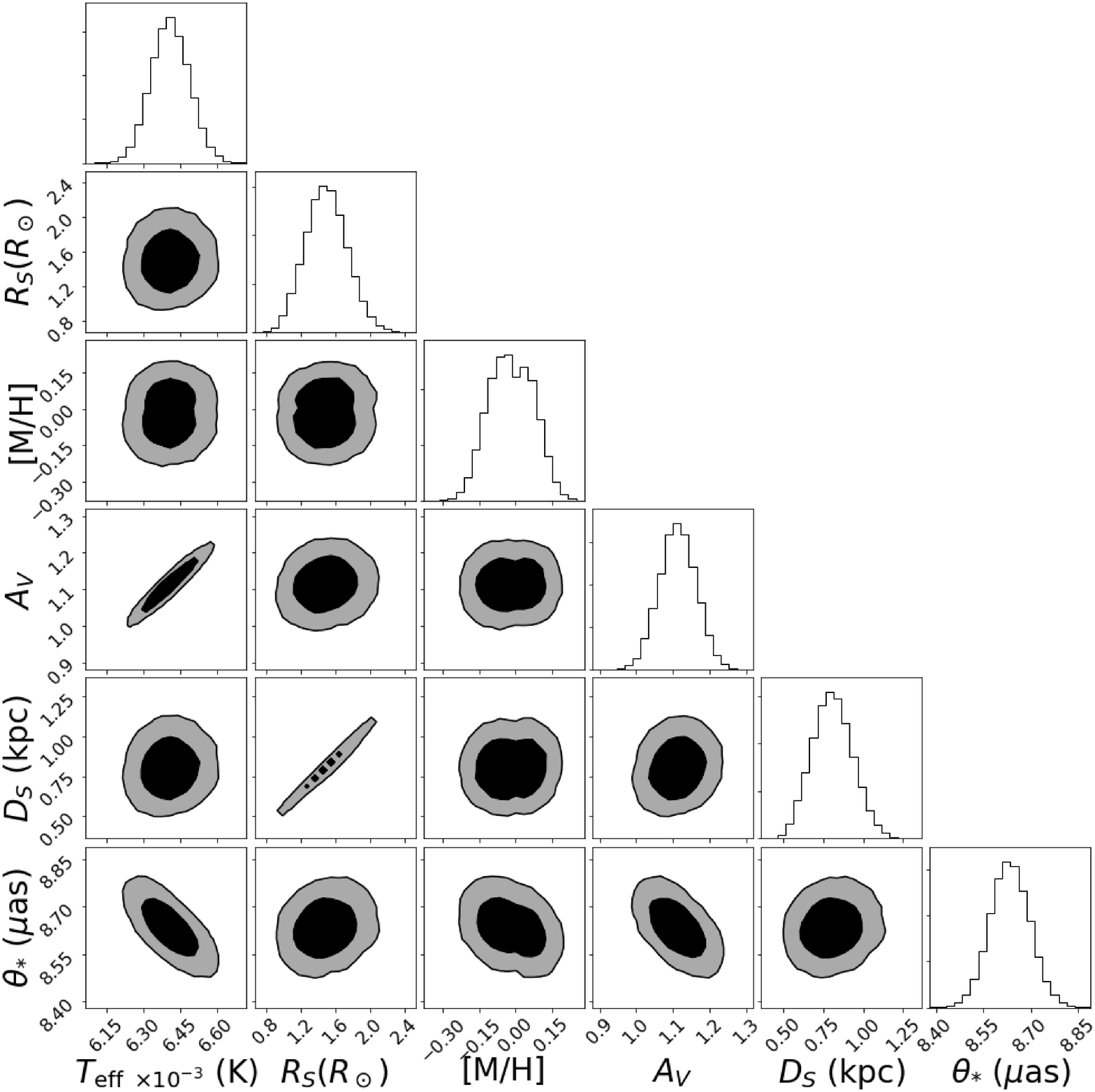}
\caption{
Corner plot for the parameters of the source star. The black and gray areas indicate the 68\% and 95\% confidence regions, respectively. Note that the bimodal feature in [M/H] centered at [M/H]=0 is an artifact due to the discreteness of the theoretical models we adopt.
\label{fig:corner_source}}
\end{center}
\end{figure*}

The derived median value and 1$\sigma$ uncertainties of the parameters are reported in Table \ref{tbl:source}, and the posterior distributions are plotted in Figure \ref{fig:corner_source}. \red{We derive the distance and angular radius of the source star to be $D_S = 800 \pm 130$~pc and $\theta_* = 8.63 \pm 0.06$~$\mu$as, respectively, which are well consistent with the previous estimations of $D_S =$~700-800~pc \citep{2018MNRAS.476.2962N} and $\theta_* = 9 \pm 0.9$~$\mu$as \citep{2019ApJ...871...70D}. }

\begin{deluxetable}{ccc}
\tablecaption{Properties of the Source Star.
\label{tbl:source}}
\tabletypesize{\footnotesize}
\tablehead{
\colhead{Parameter} & \colhead{Unit} & \colhead{Value}
}
\startdata
$g'$& mag & $14.559 \pm 0.010$\\
$V$& mag & $14.151 \pm 0.005$\\
$r'$& mag & $13.847 \pm 0.008$\\
$i'$& mag & $13.556 \pm 0.010$\\
$z'$& mag & $13.376 \pm 0.009$\\
$K_s$& mag & $11.990 \pm 0.012$\\
Effective temperature, $T_\mathrm{eff}$ &  K & $6407\ ^{+81}_{-78}$\\
Radius, $R_S$ &  $R_\odot$  & $1.49 \pm 0.25$\\
Metallicity, [M/H] & dex & $-0.02 \pm 0.10$\\
Extinction, $A_{V,S}$ & & $1.11 \pm 0.05$\\
Angular radius, $\theta_*$ & $\mu$as & $8.63 \pm 0.06$\\
Distance, $D_S$ & $10^2$ pc& $8.0 \pm 1.3$\\
\enddata
\end{deluxetable}

\section{Physical Parameters of the Lens System}
\label{sec:lens_properties}

\subsection{\red{Constraint from the Microlensing Model}}
\label{sec:from_microlens_model}

\begin{figure*}
\begin{center}
\includegraphics[width=12cm]{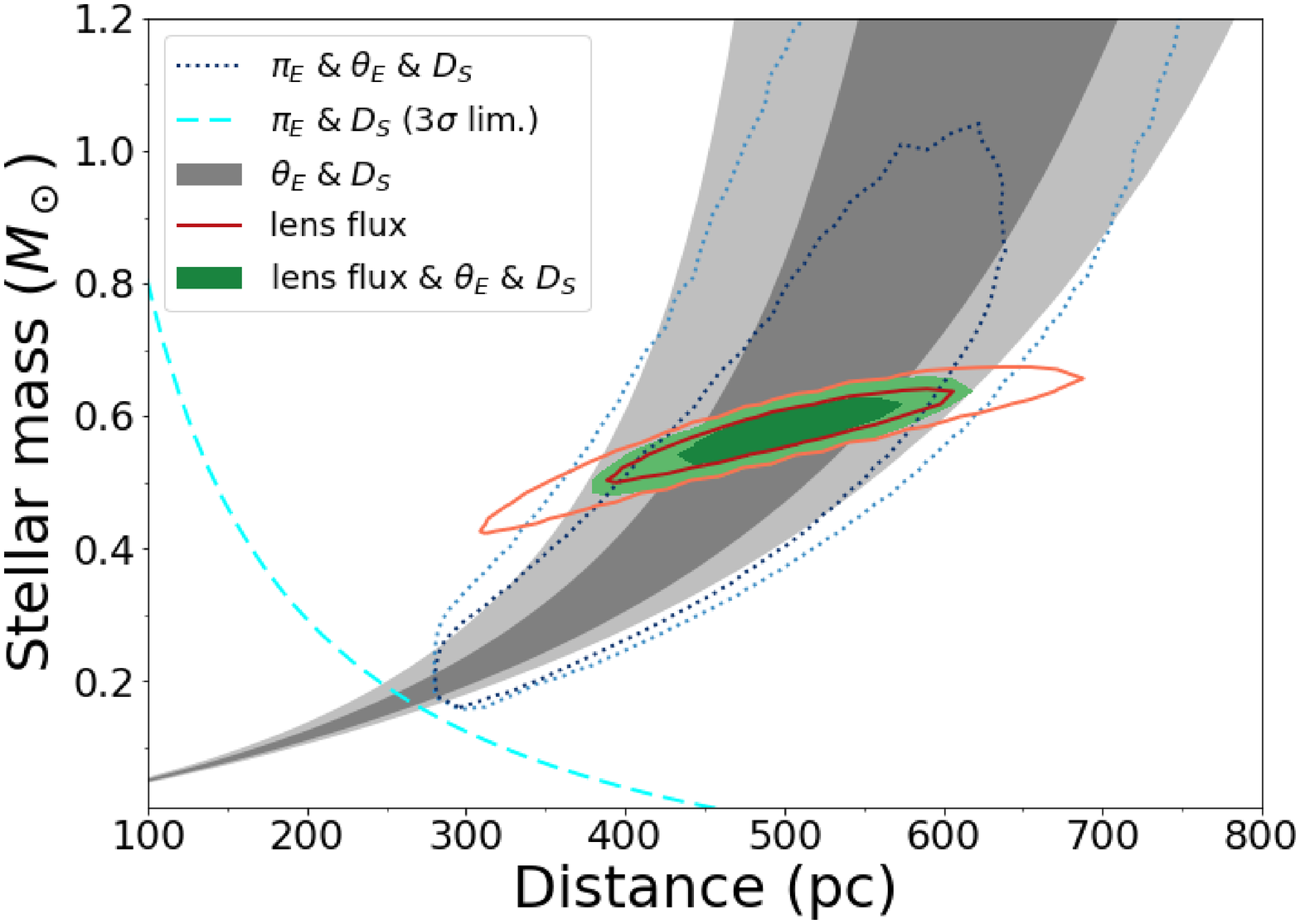}
\caption{Posterior distributions of the mass and distance of the lens star.
Blue dotted contours, gray shaded regions, red solid contours, and green shaded regions indicate the constraints calculated from $\pi_{\rm E}$ and $\theta_{\rm E}$ and $D_S$; $\theta_{\rm E}$ and $D_S$; lens flux; and the combination of lens flux and $\theta_{\rm E}$ and $D_S$, respectively. In each case, dark (inner) and light (outer) colored lines or shaded regions represent 68\% and 95\% confidence regions, respectively. The cyan dashed line indicates a lower limit given by the 3$\sigma$ upper limit of $\pi_{\rm E}$ and 3$\sigma$ lower limit of $D_S$.
\label{fig:ML_vs_DL}}
\end{center}
\end{figure*}

If $\theta_{\rm E}$, $\pi_{\rm E}$, and $D_S$ are all measured, one can solve for the total mass, $M_L$, and distance, $D_L$, of the lens system using the following formulae:
\begin{eqnarray}
\label{eq:M_L} M_L &=&  \frac{\theta_{\rm E}}{\kappa \pi_{\rm E}},\\
\label{eq:D_L} D_L &=& \frac{AU}{\pi_{\rm E} \theta_{\rm E} + \pi_S},
\end{eqnarray}
where $\pi_S \equiv AU/D_S$. The masses of the host star and planet of the lens system are then calculated as $M_{L1} = 1/(1 + q) M_L$ and $M_{L2} = q/(1+q)  M_L$, respectively, and the projected separation between the two lens components is derived by $a_{\rm proj} = s \theta_{\rm E} D_L$. The median and 1$\sigma$ uncertainties of these parameters derived from the light-curve analysis using the $\theta_{\rm E}$ and $\Phi_\pi$ priors (Section \ref{sec:parallax_with_thetaE_Phipi}) are reported in Table \ref{tbl:param_lens}, and the 68\% and 95\% confidence intervals of $M_{L1}$ and $D_L$ are shown by blue dotted lines in Figure \ref{fig:ML_vs_DL}.

However, as discussed in Section \ref{sec:parallax_without_prior}, the detection of $\pi_{\rm E}$ is marginal, and the signal is as weak as the level of systematics. Therefore, it is conservative not to rely on the $\pi_{\rm E}$ measurement to derive the lens parameters. In this case, we cannot uniquely solve for $M_{L1}$ and $D_L$ but can only draw a relation between them, as shown by the gray shaded region in Figure \ref{fig:ML_vs_DL}.

\begin{deluxetable*}{cccccc}
\tablecaption{Physical Parameters of the Lens System
\label{tbl:param_lens}}
\tablehead{
\colhead{Parameter} & \colhead{Unit} & \cite{2018MNRAS.476.2962N} & \colhead{$\pi_{\rm E}$ and $\theta_{\rm E}$ and $D_S$} & \colhead{Lens Flux} & \colhead{Lens Flux and $\theta_{\rm E}$ and $D_S$}
}
\startdata
Distance, $D_L$ & pc & $\sim$380 & $511\ ^{+101}_{-80}$& $507 \pm  74$ & $\red{505 \pm 47}$ \\
Stellar mass, $M_{L1}$ & $M_\odot$ & $0.25 \pm 0.18$ & $0.64\ ^{+0.38}_{-0.19}$ & $0.590\ ^{+0.042}_{-0.051}$ & \red{$0.586 \pm 0.033$} \\
Stellar radius, $R_{L1}$ & $R_\odot$ & -- & -- & $0.599\ ^{+0.056}_{-0.061}$ &-- \\
Extinction, $A_{V,L}$ & & -- & -- & $0.95 \pm 0.11$  &-- \\ 
Metallicity, [Fe/H] & dex & -- & -- & $-0.05\ \pm 0.20$ & -- \\
Absolute $K_s$ magnitude, $M_{K_s}$ & mag & -- & -- & $5.05\ ^{+0.33}_{-0.28}$ & -- \\
Planetary mass, $M_{L2}$ & $M_\oplus$ & $9.2 \pm 6.6$ & $21.8\ ^{+12.9}_{-6.5}$ & $20.0 \pm 2.3$ & \red{$20.0\ \pm 2.0$} \\
\red{Projected separation, $a_{\rm proj}$ (model $a$)} & au & $\sim$0.5 & \red{$0.89\ ^{+0.18}_{-0.14}$}  & $0.89 \pm 0.13$ & \red{$0.88 \pm 0.08$}\\
Projected separation, $a_{\rm proj}$ \red{(model $b$)} & au & $\sim$0.5 & $0.95\ ^{+0.19}_{-0.15}$  & $0.95 \pm 0.14$ & \red{$0.94 \pm 0.09$} \\
Semi-major axis, $a_{\rm circ}$ \tablenotemark{a} & au & -- & \red{$1.12\ ^{+0.66}_{-0.25}$} & $1.10\ ^{+0.63}_{-0.22}$ & \red{$1.08\ ^{+0.62}_{-0.18}$} \\
\enddata
Note.
\tablenotetext{a}{Calculated by merging the posteriors of models $a$ and $b$.}
\end{deluxetable*}

\subsection{From the Lens Brightness}

\subsubsection{Probabilities of Flux Contamination}
\label{sec:contami}

From the light-curve fitting, we clearly detect the blending flux in the photometric aperture, $F_b$, in the optical and near-infrared bands from $g$ through $K_s$. 
The $F_b$ values in the $g$, $r$, $i$, $z_s$, $V$, and $K_s$ bands are converted to the SDSS $g'$, $r'$, $i'$, $z'$, Johnson $V$, and 2MASS $K_s$ magnitudes, respectively, as listed in Table \ref{tbl:lens_mag}.

Generally, there are four possible sources that could contribute to the blending flux: the lens host, unrelated ambient stars, a companion to the source star, and a companion to the lens star.
In the case of this event, however, the contribution from the ambient stars is negligible because \red{the Keck AO image shows} no stars with $K_s < 21~{\rm mag}$ in the sky area of $8'' \times 8''$ other than the target.

Following the method developed by \cite{2017AJ....154....3K} and Koshimoto, N. et al. (2019 in preparation), we calculate the probabilities of all possible combinations of the other three sources that explain the observed blending flux, the Keck contrast curve (Figure \ref{fig:AO}),  and the fact that the light curve shows no significant signal of a companion.
In the calculation, we use the observed source and blending fluxes in the $V$, $I$, and $K_s$ bands, where the fluxes in the $I$ band are converted from those of $i'$- and $z'$-band magnitudes.
We do not include stellar remnants.
Using the posterior distribution from the MCMC calculation with the $\theta_{\rm E}$ prior \red{and the upper limit on $\pi_{\rm E}$ ($<$1.1)}, we calculate the probability distributions of the fraction of the lens flux to the total blending flux, $f_L \equiv F_L / F_b$, where $F_L$ is the flux from the lens star.
We find that the \red{probability of $f_L > 0.90$ is 91.8\%, 
which indicates that most of the blending flux most likely comes from the lens star.}
In the rest of the paper, we simply assume that the blending flux comes solely from the lens star.
We note that the mass and distance of the lens star derived from the blending flux under the above assumption are well consistent with the constraint from $\theta_{\rm E}$ and $D_S$ (Section \ref{sec:from_microlens_model}), supporting this assumption. There is still a small probability (8.2\%) that more than 10\% of the blending flux comes from a companion to the lens or source stars, which can be tested by direct imaging or spectroscopy of the lens star in the future.

\subsubsection{Estimation of the Mass and Distance}
\label{sec:from_lens_flux}

With the assumption that the blending flux comes solely from the lens star, we can estimate the mass and distance of the lens star using the multiband blending flux.
From an initial investigation, we find that the observed magnitudes and colors of the lens star are consistent with a main-sequence low-mass star. In estimation of the mass of low-mass stars, it is generally more reliable to use an empirical way rather than theoretical models \citep[e.g.,][]{2012ApJ...757..112B}. Therefore, to estimate a more accurate mass of the lens star, we adopt a mass-luminosity relation of \cite{2019ApJ...871...63M}, which is a fully empirical and precise (2--3\% error on mass) mass-absolute-$K_s$ relation for stars with a mass between  0.075 $M_\odot$ and 0.7 $M_\odot$, derived  based on the apparent $K_s$ magnitudes, trigonometric parallaxes, and dynamically determined masses of visual binaries.  However, \cite{2019ApJ...871...63M} provided the relation only in the $K_s$ band, with which alone the mass and distance of the lens star are degenerate for a given apparent $K_s$-band magnitude.

We therefore first solve for the distance and absolute $K_s$ magnitude, $M_{K_s}$, from the apparent $g'$-, $r'$-, $V$-, $i'$-, $z'$-, and $K_s$-band magnitudes of the host star using empirical radius-metallicity-luminosity relations from \cite{2015ApJ...804...64M}. 
They provided the relations based on spectroscopically measured effective temperatures, bolometric fluxes, metallicities, and trigonometric  parallaxes of nearby $M$--$K$ dwarfs in the form of 
\begin{eqnarray}
\label{eq:Rs}
R_* = \sum_i^n a_i M_\lambda^i  \times (1 + f \mathrm{[Fe/H]}),
\end{eqnarray}
where $R_*$ is the stellar radius, $M_\lambda$ is the absolute magnitude in the $\lambda$ band,  and $a_i$ and $f$ are coefficients. Because only the coefficients for the $K_s$ band are provided in their paper, while they also collected apparent magnitudes in other bands, including the $g'$, $r'$, $V$, $i'$, and $z'$ bands, we derive the coefficients for these additional bands from the data sets of \cite{2015ApJ...804...64M} in the same way as they did for the $K_s$ band (see the Appendix). We fit the observed magnitudes of the host star with a prediction calculated by 
\begin{eqnarray}
m_{\lambda,\mathrm{calc}} = M_\lambda + 5 \log_{10} (D_L / 10 \mathrm{pc}) + A_{\lambda,L},
\end{eqnarray}
where $\lambda$ is a given band, $D_L$ is the distance to the lens in pc, and $A_{\lambda,L} \equiv A_{V,L} \times A_\lambda/A_V$ is the extinction to the lens in the $\lambda$ band. Note that $M_\lambda$ is tied with the radius, $R_{L1}$, and metallicity, [Fe/H], of the lens star via Equation (\ref{eq:Rs}). Here we adopt $A_\lambda/A_V$ = (1.223, 1.011, 0.880, 0.676, 0.485, 0.117) for $\lambda$=($g'$, $r'$, $V$, $i'$, $z'$, $K_s$), calculated assuming $R_V = 3.1$.

We perform MCMC to derive the posterior distributions of $D_L$, $R_{L1}$, [Fe/H], and $A_{V,L}$ using the {\tt emcee} code \citep{2013PASP..125..306F}. In this calculation, we evaluate the following $\chi^2$ value:
\begin{eqnarray}
\chi^2 &=& \sum_{\lambda=\{g',r',V,i',z',K_s\}} \frac{(m_{\lambda,\mathrm{obs}} - m_{\lambda,\mathrm{calc}})^2}{\sigma_{m_{\lambda,\mathrm{obs}}}^2} \nonumber\\
&& + \frac{ (\mathrm{[Fe/H]} - \mathrm{[Fe/H]}_\mathrm{prior})^2 }{ \sigma_{\mathrm{[Fe/H]}_\mathrm{prior}}^2 }\nonumber\\
&& + \frac{ (A_{V,L} - A_{V,L,\mathrm{prior}})^2 }{ \sigma_{A_{V,L,\mathrm{prior}}}^2 },
\end{eqnarray}
where $m_{\lambda,\mathrm{obs}}$ and $\sigma_{m_{\lambda,\mathrm{obs}}}$ are the observed magnitude and its 1$\sigma$ uncertainty in the $\lambda$ band, respectively; $\mathrm{[Fe/H]}_\mathrm{prior}$ is a prior for [Fe/H]; and $A_{V,L,\mathrm{prior}}$ is a prior for $A_{V,L}$. Because our data alone do not put any meaningful constraint on [Fe/H], we impose a gaussian prior with [Fe/H] = -0.05 $\pm$ 0.20, which is from the metallicity distribution of a nearby M dwarf sample \citep{2014ApJ...791...54G}. We also take advantage of the extinction measurements of {\it Gaia} by applying Equation (\ref{eq:Av}) to $A_{V,L,\mathrm{prior}}$ and 0.10 to $\sigma_{A_{V,L,\mathrm{prior}}}$ in the same way as for $A_{V,S}$.
The derived posterior distributions of $R_{L1}$ and [Fe/H] are used to calculate the probability distribution of $M_{K_s}$ via Equation (\ref{eq:Rs}), which then gives the probability distribution of $M_{L1}$ via the mass-luminosity relation of \cite{2019ApJ...871...63M} (Equation (2) of their paper where $n$=5 is applied).

\begin{figure*}[h]
\begin{center}
\includegraphics[width=12cm]{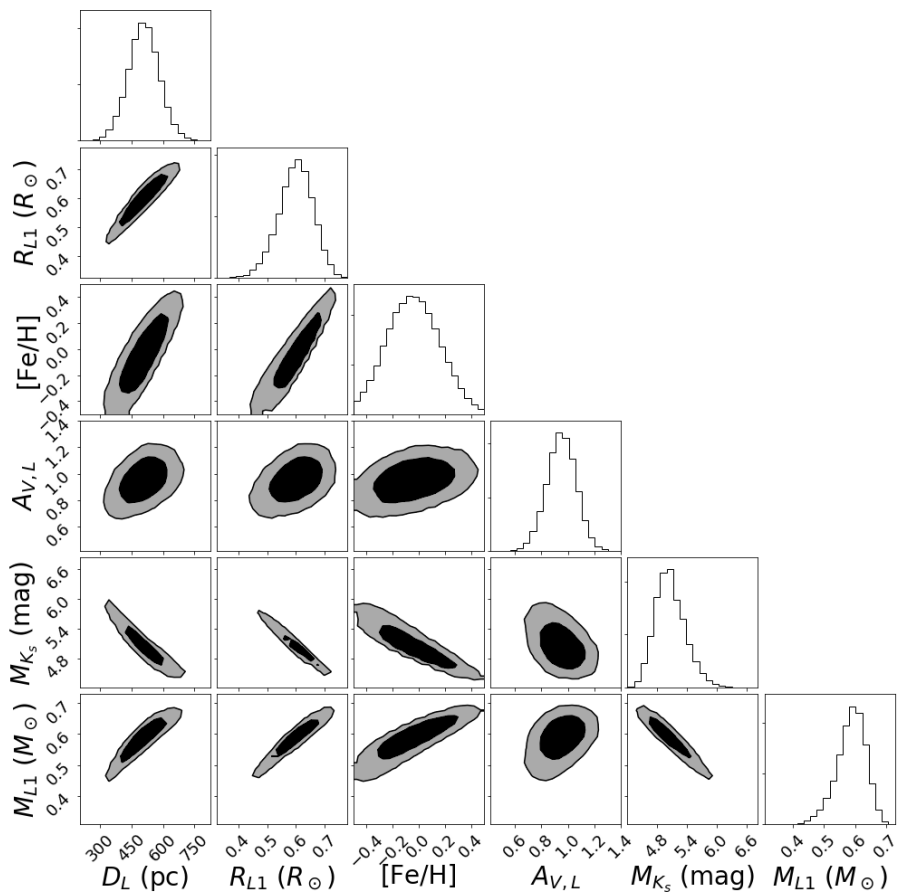}
\caption{
Corner plot for the parameters of the lens star derived from the lens brightness. The \redred{black} and \redred{gray} areas indicate the 68\% and 95\% confidence regions, respectively.
\label{fig:corner_lens}}
\end{center}
\end{figure*}

The derived median value and 1$\sigma$ uncertainties of the parameters are presented in Table \ref{tbl:param_lens}, and the posterior distributions of the parameters are plotted in Figure \ref{fig:corner_lens}. The posterior distribution between $D_L$ and $M_{L1}$ is also plotted in red in Figure \ref{fig:ML_vs_DL}. The derived $D_L$ and $M_{L1}$ are well consistent with the constraints from the microlensing model (blue dotted contours and gray shaded region in Figure \ref{fig:ML_vs_DL}), while $M_{L1}$ is much better constrained by the lens flux.

\begin{deluxetable}{cc}
\tablecaption{Calibrated Magnitudes of the Blending Flux.
\label{tbl:lens_mag}}
\tablehead{
\colhead{Band} & \colhead{Magnitude}
}
\startdata
$g'$& $19.088 \pm 0.337$\\
$V$& $17.760 \pm 0.110$\\
$r'$& $17.305 \pm 0.122$\\
$i'$& $16.382 \pm 0.068$\\
$z'$& $15.872 \pm 0.051$\\
$K_s$& $13.728 \pm 0.027$\\
\enddata
\end{deluxetable}

\subsection{Combined Solution}
\label{sec:combined}

We derive the final values of $M_{L1}$ and $D_L$ by combining the two posterior distributions, one is from \red{the microlens model (Section \ref{sec:from_microlens_model})} and the other from the lens brightness (Section \ref{sec:from_lens_flux}). \red{For the microlens model, we use the posterior distribution of the $M_{L1}$-$D_L$ relation derived from $\theta_{\rm E}$ and $D_S$ instead of the posterior distribution of the $M_{L1}$ and $D_L$ solution from $\pi_{\rm E}$, $\theta_{\rm E}$, and $D_S$, because the latter one relies on the posterior distribution of $\pi_{\rm E}$, which could be affected by systematics (Section \ref{sec:parallax}). Note that the posterior distribution from the lens flux and that from the microlens model can, in principle, be correlated because  the blending flux that the former solution relies on was also derived using the microlens model. However, this effect is so small that these two distributions} can be considered to be independent.

The combined posterior distribution is shown in green in Figure \ref{fig:ML_vs_DL}. As a result, we find that $D_L = 505 \pm 47$~pc and $M_{L1} = 0.586 \pm 0.033$~$M_\odot$; thus, the host star is a late-K/early-M boundary dwarf. The planetary mass is $M_{L2} \equiv q M_{L1} = 20.0 \pm 2.0$~$M_\oplus$, which is similar to the mass of Neptune (17.2$M_\oplus$). The sky-projected separation between the planet and the host star is $a_\mathrm{proj} \equiv s \theta_{\rm E} D_L  =$ $0.88 \pm 0.08$~AU (model $a$) and $0.94 \pm 0.09$~AU (model $b$),  which is converted to the semi-major axis of $a_{\rm circ} = 1.08 ^{+0.62}_{-0.18}$~AU, where a circular orbit and random orientation are assumed and the solutions of two models (model $a$ and $b$) are merged.

\section{Discussions}
\label{sec:discussions}

\subsection{Comparison of the Planetary Location with the Snow Line}

\begin{figure*}
\gridline{\fig{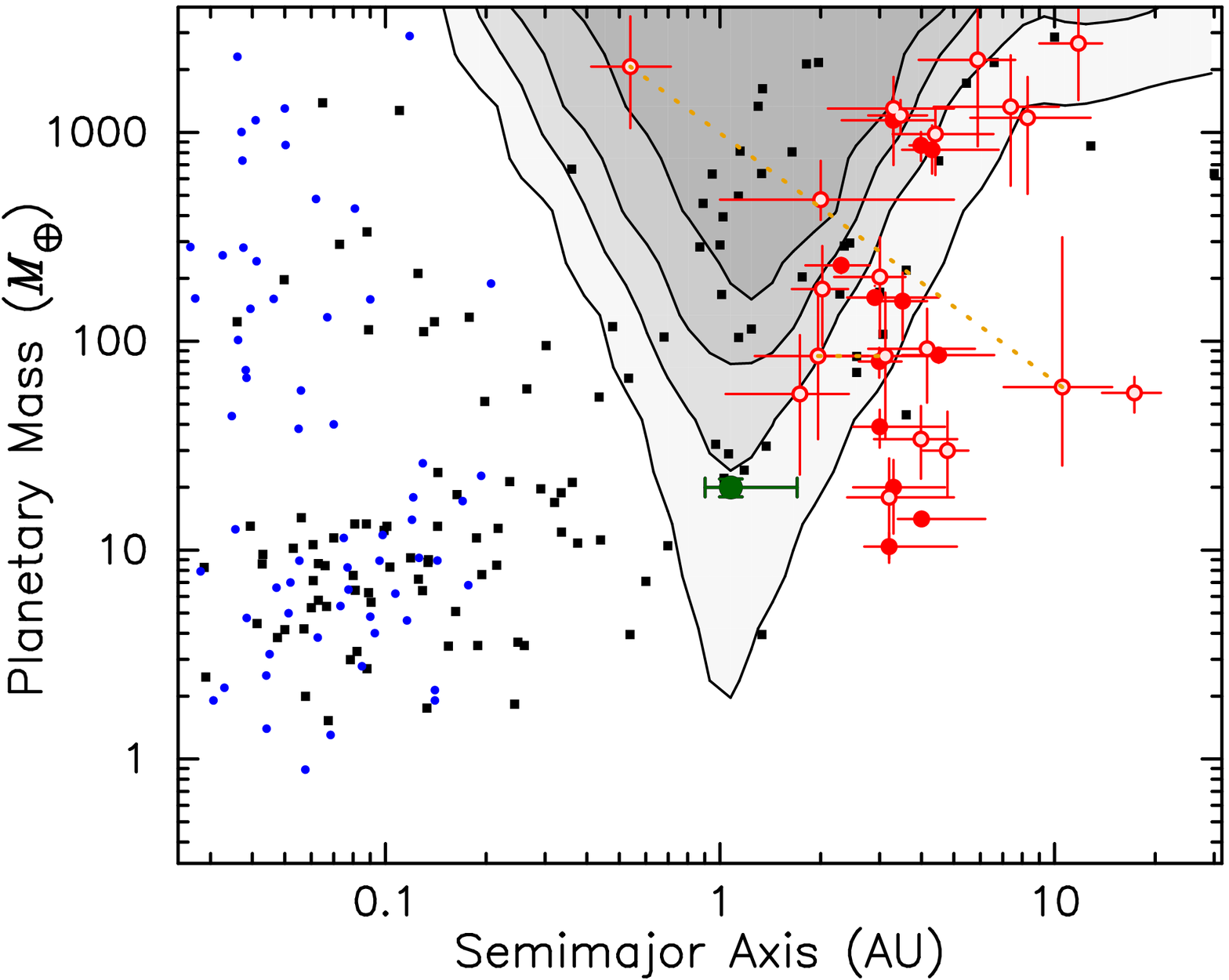}{0.4\textwidth}{(a)}
\fig{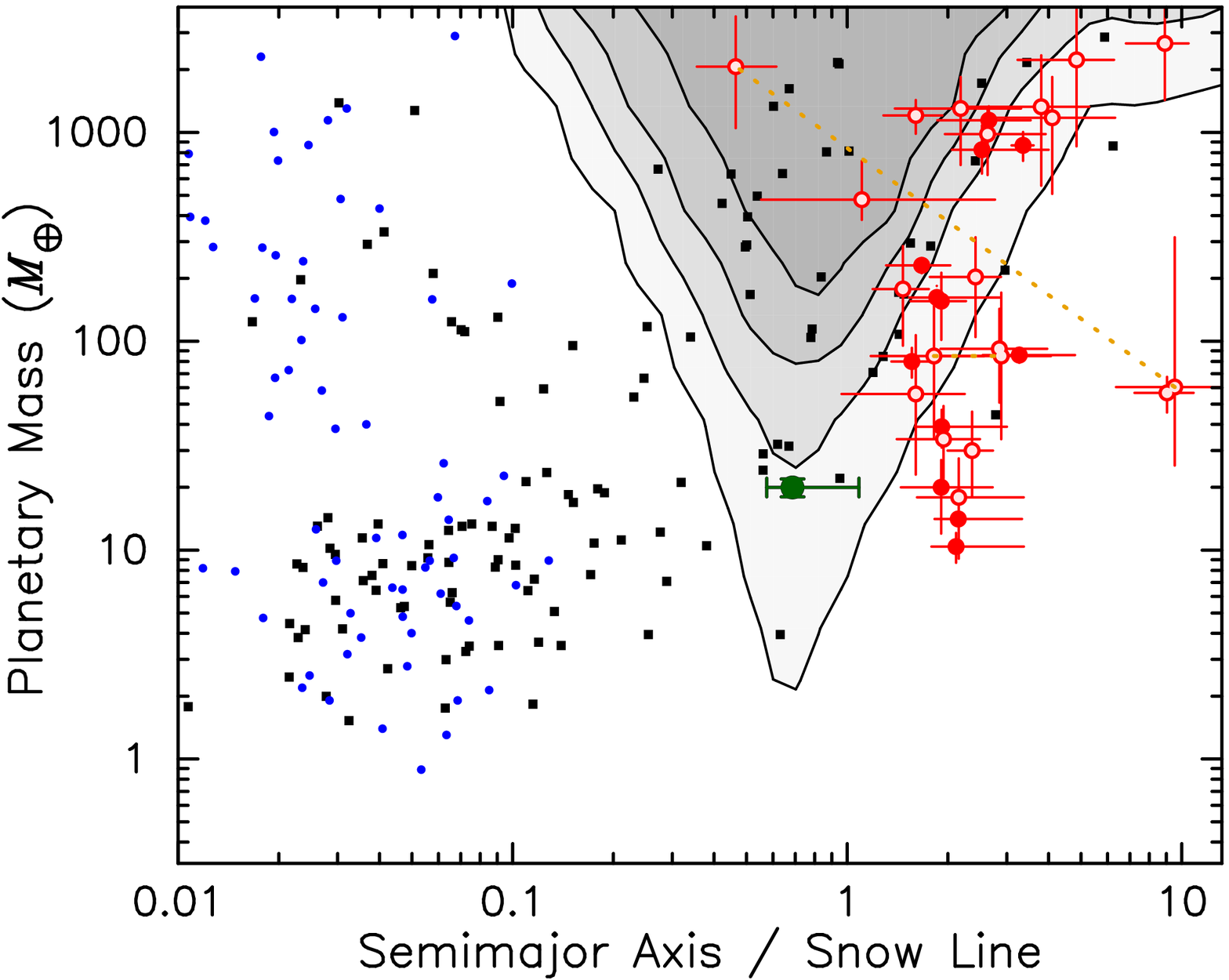}{0.4\textwidth}{(b)}}
\caption{(a) Distribution of known exoplanets in the planetary mass and semimajor axis planes for the host stars having a mass of 0.4--0.8~$M_\odot$. Data are collected mainly from http://exoplanet.eu. Black squares, blue circles, and red circles indicate the planets observed by radial velocity, transit, and microlensing, respectively. The filled and open circles of microlensing show the planets with and without direct mass constraint, respectively. Two degenerated solutions are connected by a dotted line, if applicable. Kojima-1Lb is depicted as a green circle. The contours show the planet detection efficiencies for Kojima-1 of 90\%, 70\%, 40\%, and 10\% (top to bottom). (b) Same as (a), but the $x$-axis is converted to the semimajor axis normalized by the snow-line location estimated by $a_\mathrm{snow} = 2.7 \times M_*/M_\odot$~au.
\label{fig:DE}}
\end{figure*}

Figure \ref{fig:DE} (a) shows the location of Kojima-1Lb in the plane between the mass and semimajor axis, along with the known exoplanets hosted by stars with masses similar to that of Kojima-1L (0.4--0.8~$M_\odot$). Kojima-1Lb is placed at the region where only a little has yet been surveyed by any methods due to the limitation of their sensitivity.
Several planets have been discovered in the same region with the radial velocity technique \red{\citep[e.g.,][]{2011A&A...526A.111M,2017A&A...602A..88A}}, which, however, provides only a lower limit on their masses. On the other hand, the absolute mass of Kojima-1Lb is measured with an uncertainty of only 10\%.

The orbit of Kojima-1Lb was likely comparable to the snow line at its younger age, when the planet probably formed from a protoplanetary disk.
We estimate that the snow-line location in the protoplanetary disk of Kojima-1L is $\sim$1.6~au by using the conventional formula of $a_\mathrm{snow} = 2.7 \times M_*/M_\odot$~au \citep[e.g.,][]{2008ApJ...684..663B,2010ApJ...710.1641S,2011ApJ...741...22M}, where $M_*$ is the stellar mass. This mass-linear relation can be derived by assuming that the stellar luminosity is proportional to $M_*^2$ and the protoplanetary disk is optically thin \citep{2008ApJ...684..663B}. Under this simple assumption, the present location of Kojima-1Lb is comparable to or slightly inner than the snow-line location of its youth, as shown in Figure \ref{fig:DE} (b).

\begin{figure}
\begin{center}
\includegraphics[width=8cm]{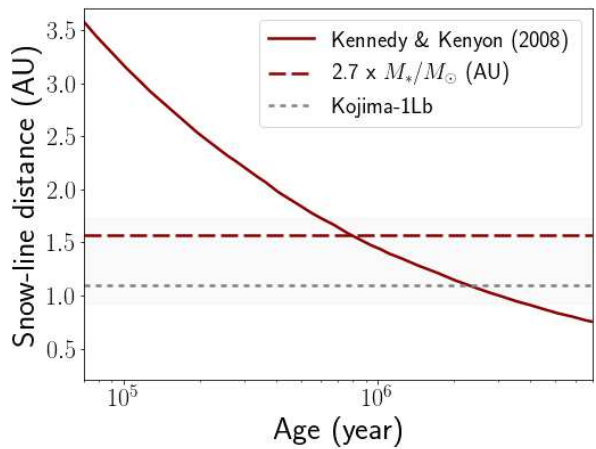}
\caption{Snow-line distance as a function of time. The solid line indicates a theoretical model for a disk of a 0.6$M_\odot$ star considering stellar irradiation and viscous accretion, extracted from Figure~1 of \cite{2008ApJ...673..502K}. The dashed line is a time-independent snow-line location for Kojima-1L calculated by $a_\mathrm{snow} = 2.7 \times M_*/M_\odot$~au. The median value and 1$\sigma$ confidence region of the semimajor axis of Kojima-1Lb are shown as a gray dotted line and light gray shaded area, respectively.
\label{fig:snowline}}
\end{center}
\end{figure}

More realistically, the snow-line distance is a function of age due to the evolution of the protoplanetary disk and stellar luminosity \citep[e.g.,][]{2006ApJ...650L.139K,2008ApJ...673..502K}. In Figure \ref{fig:snowline}, we compare the orbit of Kojima-1Lb with a theoretical prediction of the time evolution of the snow-line location at the midplane of a young disk around a 0.6 $M_\odot$ star by \cite{2008ApJ...673..502K} (extracted from Figure~1 of their paper). The model assumes stellar irradiation and viscous accretion as the sources of disk heating. According to this model, the snow-line distance monotonically decreases with time, crossing the current planet location at an age of $2.2 ^{+1.7}_{-1.6}$~Myr. This timescale is comparable to or shorter than the typical disk lifetime of low-mass stars of a few tens of Myr \citep[e.g.,][]{2012ApJ...758...31L,2015A&A...576A..52R}, indicating that the current location of Kojima-1Lb could have experienced a period when it was outside the snow line while disk gas remained.

According to the core accretion theories, it is difficult to form a planet as massive as Kojima-1Lb (20$\pm$2~$M_\oplus$) inside the snow line because of the lack of materials \citep[e.g.,][]{2005ApJ...626.1045I,2006ApJ...650L.139K}, unless the surface density of solid materials in the disk's inner region is substantially high \citep[e.g.,][]{2012ApJ...751..158H,2015A&A...578A..36O}.
On the other hand, in-situ formation of Kojima-1Lb would be possible during the period when the snow line was inside the orbit of Kojima-1Lb and the disk gas still remained. Solid materials are thought to be abundant around the snow line \citep[e.g.,][]{2002ApJ...581..666K,2017A&A...608A..92D}, which would allow the protoplanet of Kojima-1Lb to reach a mass of several $M_\oplus$ and start to accrete the surrounding gas.
Several population-synthesis studies including type I migration also predict efficient formation of Neptune-mass planets near the snow line \citep[e.g.,][]{2005ApJ...626.1045I,2009A&A...501.1161M}, while the recent result of microlensing surveys has required some modifications of these predictions, at least for the region \red{outside} a few times the snow line \citep{2018ApJ...869L..34S}.
Although it is not possible to identify the exact formation process of this specific planet, given the precise mass determination of Kojima-1Lb, this planet could be an important example toward understanding the planetary formation processes around the snow line.

\subsection{Detection Efficiency to the Planetary Signal}
\label{sec:DE}

It is interesting to consider the detection efficiency of the planetary signal in Kojima-1, 
as the sensitivity to the planet in this event could be different from typical microlensing events toward the Galactic bulge.

Assuming that the actual planet signal is absent, we calculate the detection efficiency by following the method of \cite{2000ApJ...533..378R}.
In this calculation, we use not only the data sets that are used for the light-curve fitting but also all of the other data sets listed in Table \ref{tbl:tel_list}, except for the SL data set that was identified to have systematics.
 On the other hand, we eliminate all data points after 2018 January 1 (HJD-2,450,000 = 8120), because we would have terminated our photometric follow-up campaign by the end of 2017 if the planetary signal was not detected.
As the threshold of signal detection, we adopt $\Delta \chi^2 = 100$ following \cite{2016ApJ...833..145S}, where $\Delta \chi^2$ is the $\chi^2$ difference between planetary and nonplanetary (single-lens) models. At first, the detection efficiency $\epsilon$ 
is computed as a function of $({\rm log}\,s, {\rm log}\,q)$. Next, we transform it to the physical parameter space, $({\rm log}\,a_{\rm proj}, {\rm log}\,M_{L2})$ \citep{2006MNRAS.367..669D}, where we use the well-constrained probability distribution function of $\theta_{E}$ and $M_{L1}$ instead of the Bayesian approach using a Galactic model.
The detection efficiency $\epsilon({\rm log}\,a_{\rm proj}, {\rm log}\,M_{L2})$ is further converted to $\epsilon({\rm log}\,a_{\rm 3D}, {\rm log}\,M_{L2})$ with the assumption that the planet has a circular orbit and random orientation. 

The calculated detection efficiency is plotted by contours in Figure \ref{fig:DE} (a). We also calculate the detection efficiency as a function of ${\rm log}\,(a_{\rm 3D}/a_{\rm snow})$ and ${\rm log}\,M_{L2}$, where $a_{\rm snow} = 2.7 \times (M_*/M_{\odot}) \rm au$, as shown in Figure \ref{fig:DE} (b).
The planet sensitivity of Kojima-1 has its peak around 1--1.4~au, or 0.7--1.0 times the snow-line distance.
This region is a few times interior to the region where the majority of microlensing planets have been discovered, reflected by the fact that the distance to the source star of Kojima-1 is $\sim 10$ times closer to us than those of the other microlensing events.

On the other hand, the detection efficiency of Kojima-1Lb is calculated to be only $\sim$35\%.
Here we remind the reader that the Kojima-1 event was not discovered by a systematic microlensing survey but was unexpectedly discovered during a nova search conducted by an amateur astronomer. Only one such event was previously known \citep[the so-called Tago event;][]{2007ApJ...670..423F,2008ApJ...677.1268G}, but in that case, no planetary signal was detected.
Therefore, although it is too early to argue statistically, the discovery of this low detection efficiency planet may imply that Neptunes are common rather than rare in this orbital region.
This result is consistent with the recent findings with the transit and radial velocity techniques that Neptunes are at least as common as \citep{2019AJ....157..218K} or more common than \citep{2019AJ....157..248H,2019arXiv190604644T} Jupiters at large orbits comparable to the snow line.

\subsection{Capabilities of Future Follow-up Observations}

Unlike many of the other microlensing planetary systems, Kojima-1L offers valuable opportunities to follow up in various ways thanks to its closeness to the Earth.
First, the geocentric source-lens relative proper motion is estimated to be $\mu_\mathrm{geo} = 25.34 \pm 0.44$~mas~yr$^{-1}$, enabling us to spatially separate the source and lens stars in $\sim$2 yr from the event using ground-based AO instruments (e.g., Keck/NIRC2) or space-based telescopes (e.g., {\it Hubble Space Telescope}). By resolving the two stars, one can confirm the relative proper motion (including its direction) and the brightness of the host star in an independent way \citep[e.g.,][]{2015ApJ...808..170B,2015ApJ...808..169B,2018AJ....156..289B}. 

Second, the host star is as bright as $K_s=13.7$, which is the brightest among all known microlensing planetary systems followed by OGLE-2018-BLG-0740L \citep{2019arXiv190500155H}, allowing spectroscopic characterizations of the host star. Low- or mid-resolution spectroscopy in the near-infrared is feasible with a $>$4~m class telescope, ideally with an AO instrument to reduce the contamination flux from the background source star. Such an observation will provide fundamental spectroscopic information on the host star, such as temperature, metallicity, and kinematics in the Galaxy. Furthermore, it is possible to search for additional inner and/or more massive planets with the radial velocity technique using an 8~m class telescope equipped with an AO-guided, near-infrared, high-dispersion spectrograph, such as Subaru/IRD. Knowing planetary multiplicity is of particular importance in understanding the formation and dynamical evolution of this planetary system. Finally, Kojima-1Lb would induce a radial velocity on the host star with an amplitude of $\sim$2.2 $\sin i$~ms$^{-1}$ and a period of $\sim$1.5 yr assuming a circular orbit, where $i$ is orbital inclination. This signal will be measurable in the era of extremely large telescopes (ELTs), offering a valuable opportunity to confirm the mass and refine the orbit of this snow-line Neptune.
 
\section{Summary}
\label{sec:summary}

We conducted follow-up observations of the nearby planetary microlensing event Kojima-1 by means of seeing-limited photometry, spectroscopy, and high-resolution imaging. We found no additional planetary feature in our photometric data other than the one that was identified by \cite{2017ATel10934....1N}. From the light-curve modeling and spectroscopic analysis, we have refined the distance and angular diameter of the source star to be  $800 \pm 130$~pc and $8.63\pm0.06 \mu$as, respectively. We have also refined the microlensing model using the prior information of $\theta_{\rm E}$ and $\Phi_\pi$ from the VLTI observation by \cite{2019ApJ...871...70D}. We confirm the presence of apparent blending flux and absence of significant parallax signal reported in the literature. We find no contaminating sources in the Keck AO image and that the detected blending flux most likely comes from the lens star.  Combining all of this information, we have directly derived the physical parameters of the lens system without relying on any Galactic models, finding that the host star is a dwarf on the M/K boundary  ($\red{0.59 \pm 0.03} M_\odot$) located at $\red{500 \pm 50}$~pc and the companion is a Neptune-mass planet ($\red{20\pm 2}$~$M_\oplus$) with a semimajor axis of $\red{\sim1.1}$~au.  

The orbit of Kojima-1Lb is a few times closer to the host star than the other microlensing planets around the same type of star and is likely comparable to the snow-line distance at its youth. 
We have estimated that the detection efficiency of this planet in this event is $\sim$35\%, which \redred{may imply} that Neptunes \red{are common} around the snow line.

The host star is the brightest ($K_s = 13.7$) among all of the microlensing planetary systems, providing us a great opportunity not only to spectroscopically characterize the host star but also to confirm the mass and refine the orbit of this planet with the radial velocity technique in the near future.

\acknowledgments

We thank the anonymous referee for a lot of thoughtful comments.
A.F. thanks T. Kimura and H. Kawahara for meaningful discussions on the formation and abundance of Neptunes around the snow line. A.F. also thanks A. Nucita and A. Mann for kindly providing data used in their papers.

This article is based on observations made with the MuSCAT2 instrument, developed by ABC, at Telescopio Carlos S\'anchez, operated on the island of Tenerife by the IAC in the Spanish Observatorio del Teide.
We acknowledge ISAS/JAXA for the use of its facility through the inter-university research system.
A.Y. is grateful to Mizuki Isogai, Akira Arai, and Hideyo Kawakita 
for their technical support on observations with the Araki telescope. 
D.S. acknowledges The Open University for the use of the COAST telescope. \red{A.F. acknowledges the MOA collaboration/Osaka University for the use of the computing cluster.}

This work was partly supported by JSPS KAKENHI grant Nos. JP25870893, JP16K17660, JP17H02871, JP17H04574, JP18H01265, and JP18H05439; MEXT KAKENHI grant Nos. JP17H06362 and JP23103004; and JST PRESTO grant No. JPMJPR1775.
This work was also partially supported by Optical and Near-Infrared Astronomy Inter-University Cooperation Program of the MEXT of Japan and the JSPS and NSF under the JSPS-NSF Partnerships for International Research and Education.
This work was partly financed by the Spanish Ministry of Economics and Competitiveness through grants ESP2013-48391-C4-2-R and AYA2015-69350-C3-2-P.
Y.T. acknowledges the support of DFG priority program SPP 1992 ``Exploring the Diversity of Extrasolar Planets'' (WA 1047/11-1). S.Sh. acknowledges the support from grants APVV-15-0458 and VEGA 2/0008/17.

\section*{Appendix}

To complement Table 1 of \cite{2015ApJ...804...64M}, we calculate the coefficients of the radius-metallicity-luminosity relation for other bands than $K_s$ band using the same data set used by \cite{2015ApJ...804...64M}. They made public a table that includes synthetic apparent magnitudes in various bands (calculated from cataloged magnitudes and low-resolution spectra) and stellar radius (estimated from the observed bolometric flux and effective temperature) for 183 nearby M7--K7 single stars. This table, however, lacks the information on parallax that is needed to convert the apparent magnitude to absolute magnitude, which we got from the authors by private communication. (Their parallax came from somewhere before {\it Gaia}, but we do not attempt to update them using {\it Gaia} to keep consistency.)

To derive the relation, we apply Equation (5) of their paper, that is,
\begin{eqnarray}
\nonumber
R_* = (a + bM_\lambda + cM_\lambda^2 + ..)\\ 
 \times (1 + f\rm{[Fe/H]}),
\end{eqnarray}
where $R_*$ is the stellar radius, $M_\lambda$ is the absolute magnitude in band $\lambda$, [Fe/H] is the metallicity, and $a$, $b$, $c$, .., $f$ are coefficients. We choose the polynomial order for  $M_\lambda$ such that the best-fit BIC value \citep{1978Schwarz} is minimized. We derive the coefficients for the $g'$, $r'$, $i'$, $z'$, and $V$ bands, as well as for the $K_s$ band, for completeness, as listed in Table \ref{tbl:coeffs}. 

\begin{deluxetable*}{ccccccc}
\tablecaption{Coefficients of Radius-metallicity-luminosity Relation.
\label{tbl:coeffs}}
\tablehead{
\colhead{Band} & \colhead{a} & \colhead{b} & \colhead{c} & \colhead{d} & \colhead{e} & \colhead{f}
}
\startdata
$g'$ & -4.0294 & 1.6103 & $-1.9349 \times 10^{-1}$ & $9.4899 \times10^{-3}$ & $-1.6655 \times 10^{-4}$ &  $3.2209 \times 10^{-1}$ \\
$r'$ & -2.5349 &   1.2698 &  $-1.7485 \times 10^{-1}$ & $9.6309 \times 10^{-3}$ & $-1.8821 \times 10^{-4}$ & $3.4127 \times 10^{-1}$\\
$i'$ & -3.5485 & 1.9081 & $-2.9955 \times 10^{-1}$ & $1.9070 \times 10^{-2}$ & $-4.3370 \times 10^{-4}$ & $2.5015 \times 10^{-1}$\\
$z'$& -3.9416 & 2.3156 & $-4.0010 \times 10^{-1}$ & $2.8101 \times 10^{-2}$ & $-7.0665 \times 10^{-4}$ & $1.766 \times 10^{-1}$\\
$V$ & -3.1842 & 1.4307 &  $-1.8538 \times 10^{-1}$ & $9.7067 \times 10^{-3}$ & $-1.8107 \times 10^{-4}$ & $3.3462 \times 10^{-1}$\\
$K_s$ & 1.9305 &  $-3.4665 \times 10^{-1}$ &  $1.6472 \times 10^{-2}$ &  -- & -- & $4.4889 \times 10^{-2}\ \tablenotemark{a}$\\
\enddata
\tablenotetext{a}{There is a small difference in the values between this work and \cite{2015ApJ...804...64M}, which we suspect due to round errors in [Fe/H].}
\end{deluxetable*}

\bibliographystyle{aasjournal}
\bibliography{ref_kojima}

\listofchanges
\end{document}